\documentclass{aastex62}

\graphicspath{{./}{figures/}}

\received{}
\revised{}
\accepted{}
\submitjournal{AJ}

\shorttitle{The G\,305 complex: Irregular variables.}
\shortauthors{Medina et al.}

\begin{document}

\title{The G\,305 Star-forming Region: II. Irregular variable stars.}

\correspondingauthor{Nicol\'as Medina Pe\~na}
\email{nicolas.medina@postgrado.uv.cl}

\author[0000-0002-0786-7307]{N. Medina}
\affil{Instituto de F\'isica y Astronom\'ia, Universidad de Valpara\'iso, Av. Gran Breta\~na 1111, Playa Ancha, Casilla 5030, Chile.}
\affil{Millennium Institute of Astrophysics, Nuncio Monsenor Sotero Sanz 100, Of. 104, Providencia, Santiago, Chile.}

\author[0000-0002-0786-7307]{J. Borissova}
\affil{Instituto de F\'isica y Astronom\'ia, Universidad de Valpara\'iso, Av. Gran Breta\~na 1111, Playa Ancha, Casilla 5030, Chile.}
\affil{Millennium Institute of Astrophysics, Nuncio Monsenor Sotero Sanz 100, Of. 104, Providencia, Santiago, Chile.}

\author[0000-0002-9740-9974]{R. Kurtev}
\affil{Instituto de F\'isica y Astronom\'ia, Universidad de Valpara\'iso, Av. Gran Breta\~na 1111, Playa Ancha, Casilla 5030, Chile.}
\affil{Millennium Institute of Astrophysics, Nuncio Monsenor Sotero Sanz 100, Of. 104, Providencia, Santiago, Chile.}

\author[0000-0003-3496-3772]{J. Alonso-Garc\'ia}
\affil{Centro de Astronom\'{i}a (CITEVA), Universidad de Antofagasta, Av. Angamos 601, Antofagasta, Chile.}
\affil{Millennium Institute of Astrophysics, Nuncio Monsenor Sotero Sanz 100, Of. 104, Providencia, Santiago, Chile.}

\author[0000-0001-8600-4798]{Carlos G. Rom\'an-Z\'u\~niga}
\affiliation{Universidad Nacional Aut\'onoma de M\'exico, Instituto de Astronom\'ia, AP 106, Ensenada 22800, BC, Mexico}

\author[0000-0001-7868-7031]{A. Bayo}
\affiliation{Instituto de F\'isica y Astronom\'ia, Universidad de Valpara\'iso, Av. Gran Breta\~na 1111, Playa Ancha, Casilla 5030, Chile.}
\affiliation{N\'ucleo Milenio Formaci\'on Planetaria - NPF, Universidad de Valpara\'iso, Av. Gran Breta\~na 1111, Valpara\'iso, Chile.}

\author{Marina Kounkel}
\affil{Department of Physics and Astronomy, Western Washington University, 516 High St., Bellingham, WA 98225, USA}

\author[0000-0002-1379-4204]{Alexandre Roman-Lopes}
\affiliation{Departamento de Astronom\'{\i}a, Universidad de La Serena, Av. Cisternas 1200N, La Serena, Chile} 

\author{P. W. Lucas}
\affil{Centre for Astrophysics, University of Hertfordshire, College Lane, Hatffeld, AL10 9AB, UK.}

\author[0000-0001-6914-7797]{K. R. Covey}
\affiliation{Department of Physics and Astronomy, Western Washington University, Bellingham, USA}

\author[0000-0003-3459-2270]{Francisco F\'{o}rster}
\affiliation{Center for Mathematical Modeling, University of Chile, AFB170001, Chile}
\affil{Millennium Institute of Astrophysics (MAS), Santiago, Chile.}

\author{Dante Minniti}
\affiliation{Departamento de F\'isica, Facultad de Ciencias Exactas, Universidad And\'es Bello, Av. Fernandez Concha 700, Las Condes,
Santiago, Chile.}

\author{Lucia Adame}
\affiliation{Universidad Nacional Aut\'onoma de M\'exico, Instituto de Astronom\'ia, AP 106, Ensenada 22800, BC, Mexico}

\author{Jes\'{u}s Hern\'{a}ndez}
\affiliation{Universidad Nacional Aut\'onoma de M\'exico, Instituto de Astronom\'ia, AP 106, Ensenada 22800, BC, Mexico}



\begin{abstract}
We present a catalog of 167 newly discovered, irregular variables spanning a $\sim$7 deg${^2}$ area that encompasses the G\,305 star-forming complex, one of the most luminous giant H\,II regions in the Galaxy. We aim to unveil and characterize the young stellar object (YSO) population of the region by analyzing the $K_{\rm s}$-band variability and $JHK_{\rm s}$ infrared colors from the {\it VISTA Variables in the V\'ia L\'actea} (VVV) survey. Additionally, SDSS-IV APOGEE-2 infrared spectra of selected objects are analyzed. 

The sample show relatively high amplitudes ($0.661<\Delta K_{\rm S} <3.521$ mag). Most of them resemble sources with outbursts with amplitude $>1$ mag and duration longer than a few days, typically at least a year, known as {\it Eruptive Variables}. About  60\,\% are likely to be Class II/Flat/I objects. This is also confirmed by the spectral index $\alpha$ when available.

From the analysis of APOGEE-2 near-infrared spectra of sources in the region, another 122 stars are classified as YSOs, and displays some infrared variability. The measured effective temperature $T_{\rm eff}$ peak is around 4000K and they are slightly super-solar in metal abundance. The modal radial velocity is approximately $-$41 km/s. 

Combining available catalogs of YSOs in the region with our data, we investigate the spatial distributions of 700 YSOs. They are clearly concentrated within the central cavity formed by the massive clusters Danks\ 1 and 2. The calculated surface density for the entire catalog is 0.025 YSOs/pc$^{-2}$, while the central cavity contains 10 times more objects per area (0.238 YSOs/pc$^{-2}$).

\end{abstract}



\section{Introduction}\label{intro}
Since stars in formation or in very early stages of evolution undergo rapid structural changes, variations in their luminosity and colours are widely observed. Historically, various types of brightness changes have been noticed in young stellar objects (YSOs). For instance, strong variability may be due to accretion events, while smaller-scale variability may be caused by geometrical effects, such as clearing of dust, spots, and rotation \citep{Audard2014}. FU Orionis stars \citep[FUors;][]{Herbig1966} are thought to be pre-main sequence (pre-MS) stars characterized by increases in brightness by $>$ 4 magnitudes in optical wavelengths. According to~\cite{Audard2014}, only about two dozen such sources were known in 2014. EX Lupi-type objects \citep[EXors;][]{Herbig89} show similar brightness increases to FUors, but with shorter timescales. There are still problems in the interpretation of the variety of outburst intensities and timescales for those types of stars. 

This variable brightness also can be observed in several stages of the classical model of pre-MS evolution, after the collapse of a molecular cloud has begun. Due to conservation of angular momentum, the infalling matter forms a protostellar disk, through which accretion on to the central source occurs \citep{2007ARA&A..45..565M,2014prpl.conf..173L}. Numerous physical processes are involved in the evolution from a heavily obscured protostar until it reaches the main sequence. The physical origin of the variability could be a consequence of irregular intrinsic changes, such as episodic accretion~\citep{Audard2014}, variable extinction induced by their circumstellar disks (e.g.,~\citealt{Meyer1997,Cody2014, Rebull2014}), rotational modulation by cool and hot spots on the stellar surface, and the variations in accretion disk/envelope geometry. Massive YSOs, with $\cal M$ $>$ $7 \cal M_{\odot}$, likely undergo slightly different accretion and formation processes \citep{2011MNRAS.416..972D,2011ApJ...730L..33M,2021arXiv210205087F}. Observations of variability may provide some insight into these processes \citep{2017NatPh..13..276C}.

~\cite{Rice2015} determined that the amplitude of brightness variations is correlated with evolutionary class in YSOs. In other words, Class I YSOs are more variable than Class II, which are in turn more variable than Class III. Therefore YSOs at earlier evolutionary stages will be preferentially identified in variability-based searches. 

Optical variability was one of the original, defining characteristics of a YSO \citep{1945ApJ...102..168J, 1952JRASC..46..222H}. Therefore, in addition to analysis based on spectroscopy, the colour-magnitude diagram (CMD), and the colour-colour diagram (CCD), the study of photometric variability has been essential to understand the underlying physical processes. Existing time-domain surveys (e.g., the Catalina Real-time Transient Survey~\citep{Drake2009}, the Pan-STARRS survey \citep{Kaiser2002}), the near future Gaia variability releases~\citep{Eyer2019} and the Vera C. Rubin Observatory, previously known as the Large Synoptic Survey Telescope \citep[LSST;][]{2019ApJ...873..111I} provide and will continue providing both large area and time-series covering in the optical wavelength regime. 

\begin{figure}[htbp]
\begin{center}
\includegraphics[height=7cm,width=14cm]{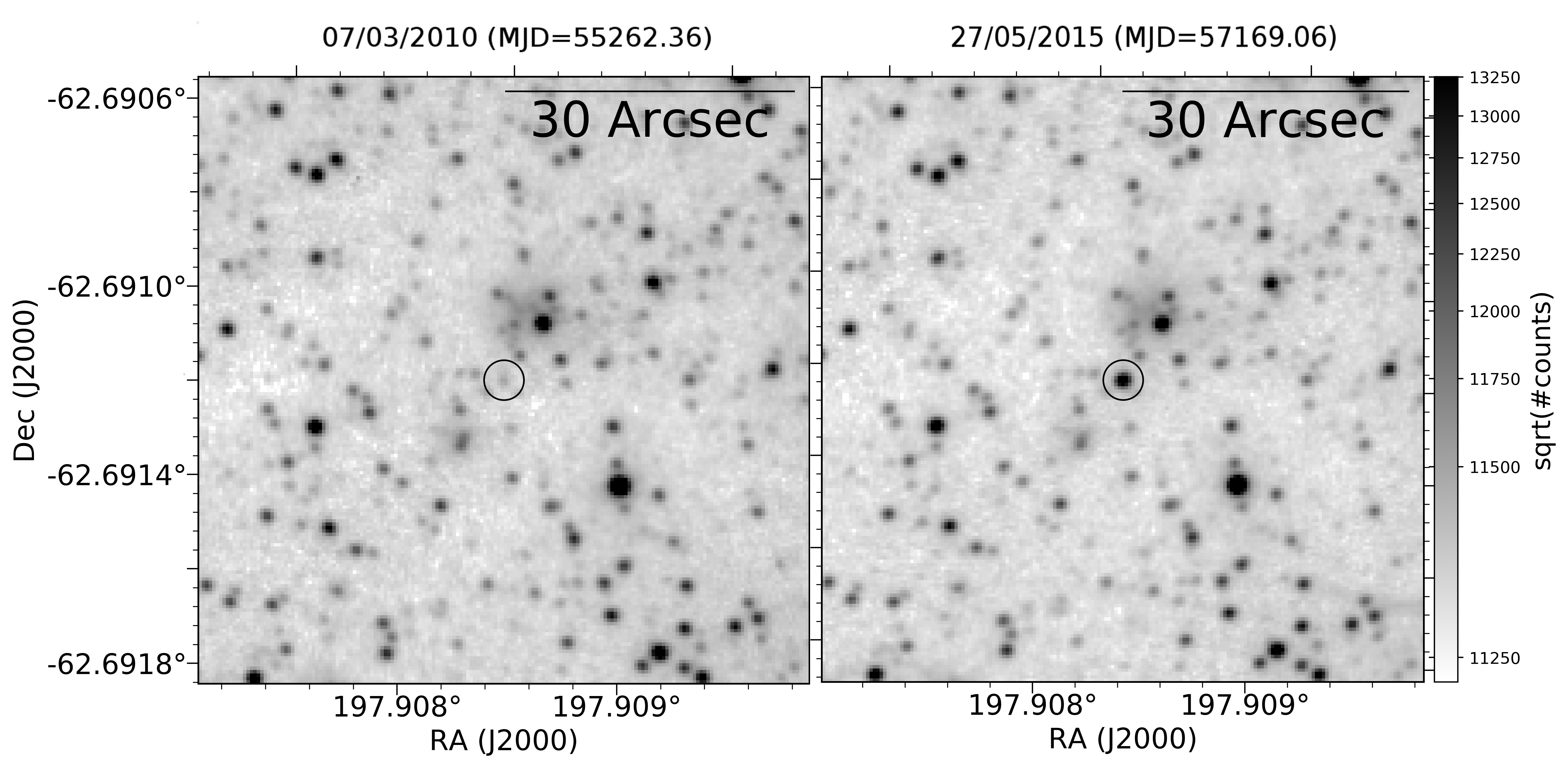}
\caption{An example of a variable star, denominated d084\_v8. Different epochs are indicated in the top of each image, the bar shows the square root of measured counts. }
\label{VariableSource}
\end{center}
\end{figure}

However, the YSOs are often located inside or adjacent to dense molecular clouds \citep{2009ApJS..181..321E} These clouds of material absorb and re-emit an important amount of the radiation from the young sources. Also, the disks and envelopes of the YSOs themselves are optically thick. Photometric surveys in the mid-infrared (MIR), such as Young Stellar Object VARiability \citep[YSOVAR;][]{2014AJ....148...92R}, or in the near-infrared (NIR) are very useful to minimize the influence of interstellar extinction. One of these NIR surveys is the 
{\it VISTA Variables in the V\'ia L\'actea} survey \citep[VVV;][]{Minniti2010,Saito2012} and its undergoing extension, the {\it VISTA Variables in the V\'ia L\'actea Extended} survey \citep[VVVX;][]{min18}. The VVV survey is fully comparable to the optical ones both in area and time-domain coverage \citep[][]{arnaboldi2007} and have been designed to catalogue more than 10$^9$ sources, where 10$^6$ are expected to be variable stars. VVV is very suitable to characterize the changes along time. Several authors reported objects with considerable amplitudes \citep[$>1$\,mag;][]{Contreras2017a} in $K$ and $K_{\rm S}$ bands, and also revealing that variability in the NIR is relatively common in YSOs. According to \cite{2012ApJ...755...65R}, out of the 30 YSOs they studied, 28 (93\%) are variable. 
Figure~\ref{VariableSource} shows an example demonstrating a significant brightness change of the variable source named d084\_v8, discovered in this work. The $K_{\rm S}$ images correspond to different epochs (marked in the top of each image).

\subsection{The G\,305 region}
 
\begin{figure}[htbp]
\begin{center}
\includegraphics[height=13cm,width=16cm]{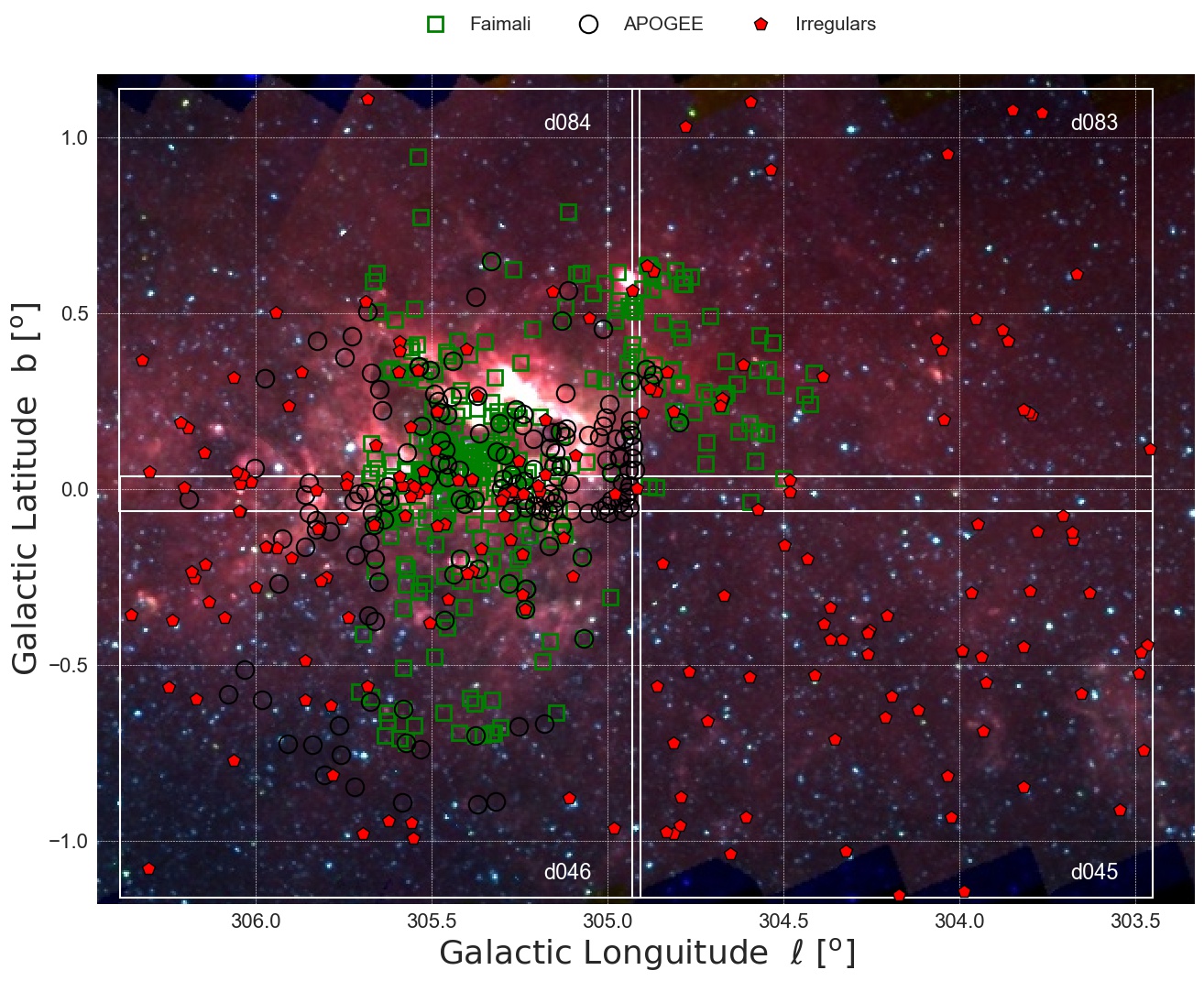}
\caption{The G\,305 star-forming complex region. The four observed tiles, named d045, d046, d083, and d084 from the VVV survey are labeled. The symbols represent different sets of YSOs and candidates discussed in this paper (see text). In the background, false three-color GLIMPSE Infrared Array Camera \citep[IRAC;][]{2004ApJS..154...10F} image of the complete field is shown, using 8.0$\mu$m (red), 4.5$\mu$m (green), and 3.6$\mu$m (blue). Galactic north is up, galactic east is to the left.}
\label{FoV}
\end{center}
\end{figure}

G\,305 (Fig.\,\ref{FoV}) is one of the most massive star-forming regions in our Galaxy, located within the Scutum-Crux arm (in galactic coordinates: $l=305.506^\circ$, $b=00.085^\circ$). Its integrated radio luminosity can be directly comparable to the most luminous giant H\,II regions of the Galaxy \citep{2004MNRAS.355..899C}. G\,305 is also a valuable laboratory for studying sequential star formation. According to \citet{Clark2004}, there are at least two different generations of young stars within the G\,305 radio complex. The first generation is associated with the massive clusters Danks\,1 and 2, which contain many massive OB and Wolf-Rayet stars. The well known Wolf-Rayet star WR48a is projected nearby. The presence of a dusty WCL star implies a minimum age of $\sim$\,3 Myr and the lack of the red supergiants in the region suggests a maximum age of $\sim$\,4-5 Myr. The second generation consists of embedded IR sources, maser emission and ultracompact H\,II regions \citep[UCHIIs;][]{hindson2012}. A significant portion of the radio emission comes from the regions which coincide with the periphery of the mid-IR nebula \citep{Clark2004}. 

This is the second paper of our series on this region. In~\cite{Borissova2019}, we used moderately high resolution spectroscopy obtained with the Apache Point Observatory Galaxy Evolution Experiment-2 \citep[APOGEE-2;][]{Majewski2017} spectrograph to study the massive stellar content in the area. We investigated 29 OB type, Wolf-Rayet, and emission-line stars, adding 18 newly identified objects to the existing catalogs. Thus, we improved the census of the hot stellar population in the region. The average spectroscopic and Gaia DR2 distances are obtained as 3.2$\pm$1.6 kpc and 3.7$\pm$1.8 kpc, respectively. The OB stars have a mean radial velocity of RV=$-41.8$ $\rm km\,s^{-1}$. Eight objects show time-series variations in the $K_{\rm S}$-bandpass with amplitudes greater than 0.5 mag. All of them, except 2MASS J13113560-6245324 (class 0/I YSO) and 2MASS J13160379-6242218 (OIf*/WN), are B type main sequence stars.

In this paper, we report a new study of the YSO population in the region. A variability search in the NIR $K_{\rm S}$-bandpass is performed. Sources are identified as variable stars by their variability indices via analysis of the time-series data of the entire VVV survey period between 2010 and 2015. Spectroscopic follow-up of a sample is performed with the APOGEE-2 spectrograph. Finally, we collated catalogs of YSOs and candidate YSOs from the literature in order to investigate their spatial distribution. 

\section{Variability search}\label{data}

The VVV is an ESO NIR public survey~\citep{Minniti2010,Saito2012} which uses the 4-meter VISTA telescope \citep{2015A&A...575A..25S} at Cerro Paranal Observatory, Chile. It was designed for mapping 562 deg$^{2}$ in the Galactic bulge and the southern disk in five NIR broad-band filters: $Z$, $Y$, $J$, $H$, and $K_{\rm S}$, which has a time coverage spanning over five years, between 2010 and mid-2015. The observations were performed with the VIRCAM NIR camera~\citep{Dalton2006}, which is an array of 16 detectors with $2048\times2048$ pixels each. The disk area is divided into 152 observing areas (1.5$\times$1.2 deg each, called tiles) and the bulge is covered with 196 tiles. Currently, the observed disk and bulge regions are extended by the ongoing VVVX ESO public survey \citep{min18}. The multi epoch observations are performed only in $K_{\rm S}$. The VVV data were our main source of information for identifying stellar objects with variable brightness.
 
According to radio observations~\citep{hindson2012}, the projected diameter of the G\,305 complex is $~$30 pc, at 3.5 kpc distance. Thus, the complete region is covered by four VVV tiles, namely d045, d046, d083 and d084, with total area of $\sim7$ deg$^{2}$. The selected regions and the total field of view (FoV) are shown in Figure~\ref{FoV}.

In order to identify and characterize variable sources in the FoV, we used the automated tool presented in \citet{Medina2018}. Briefly, each pawprint image was retrieved from the VISTA Science Archive\footnote{http://horus.roe.ac.uk/vsa} \citep[VSA;][]{Cross2012}. Then, point spread function (PSF) photometry was obtained using the $\mathtt{Dophot}$ software~\citep{Schechter1993, Alonso2012} in all available images. The calibration process on to the VISTA system was done using the aperture photometry catalogs produced by the Cambridge Astronomical Survey Unit\footnote{http://casu.ast.cam.ac.uk/} (CASU, v1.3 pipeline). We selected sources with flags -1 (``Stellar'') or -2 (``Border-line stellar'') morphological classification to perform the cross-match, using a tolerance of 0.''34 (the VISTA pixel size). The conversion factors and uncertainties were estimated using a linear fit to the $\mathtt{Dophot}$ PSF photometry vs. selected, well isolated CASU sources. A 2$\sigma$ clipping was used in the linear model to avoid the scatter caused by fainter sources ($K_{s}\sim16-17$ mag) in both photometric catalogues. We also included VVV tile images with flag ``$\mathtt{deprecated=50}$'', which indicate the atmospheric $\rm seeing > 2"$ during the observation. These images are found to be still useful for variability measurements, after careful inspection. In total, we generated and analyzed 3,570,064 $K_{\rm S}$ light curves of stars that populate the complete FoV. Table~\ref{BasicInfo} lists some basic information of the tiles, such as identification number, coordinates, and the number of available epochs. The number of sources with time-series and the number of irregular variables are also reported.

\begin{figure}[htbp]
\begin{center}
\includegraphics[height=6cm,width=\linewidth]{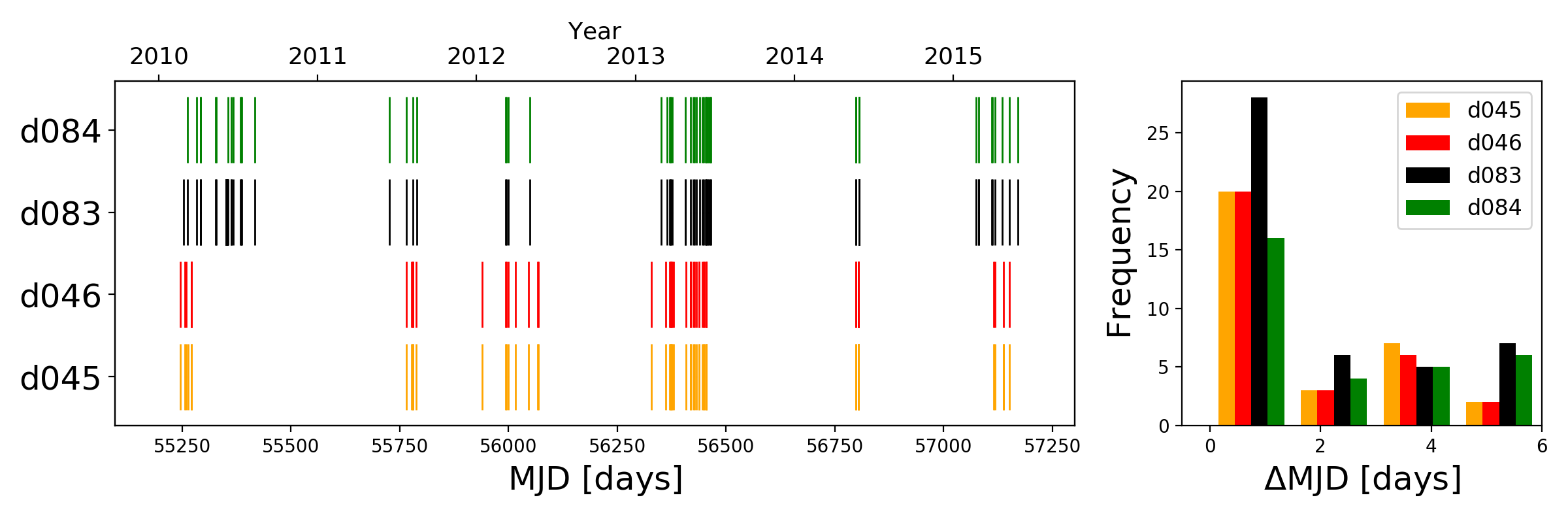}
\caption{Left panel: Cadence of the selected VVV tiles. Each photometric measurement is represented by a '$|$' symbol. The bars are thicker in places with high cadence. Right panel: Histogram of differences between consecutive observations $\Delta$MJD. The color codes represent the same tiles in each plots.}
\end{center}
\label{Cadence}
\end{figure}

As shown in Figure~\ref{Cadence}, VVV produces unevenly sampled time-series \citep{Scargle1982} that vary in sampling and cadence from tile to tile. As can be seen in the right panel, the typical time interval between observations is between 0.3 and 1.6 days for all tiles. Thus, the cadence of observations are well suited to detect short timescales variability in YSOs, related to short term accretion rate variations, episodic accretion events, rotational modulation by star spots and variable extinction~\citep{Rebull2014}. In fact, some studies have already been published regarding these topics and based on VVV data: \citet{Contreras2017a, Contreras2017b} and \citet{Lucas2017} cataloged high amplitude eruptive variable stars;~\cite{Borissova2016} searched for YSOs around young stellar clusters;~\cite{Teixeira2018} measured variability in intermediate luminosity YSOs and massive YSO candidates; \citet{Medina2018} defined a new class of slowly varying YSOs, the so called “Low Amplitude Eruptive (LAE)” variables; and  \citet{Guo2020} identified "multiple timescale variables" that are highly variable on timescales of weeks and years.

\begin{table*}[hbt]
\begin{center}
\begin{tabular}{ccccc}
\hline\hline
VVV & Central coordinates &Epochs & Sources with & Irregular \\
tile ID & ($\ell\ [^{\circ}],b\ [^{\circ}]$) &  & time-series & sources \\
\hline 
d045  & (304.2,-0.5) & 55   & 950,085         & 57            \\
d046  & (305.7,-0.5) & 54   & 1,068,806        & 66            \\
d083  & (304.2,0.5)  & 74   & 754,122         & 32            \\
d084  & (305.7,0.5)  & 59   & 797,051         & 41            \\ \hline
\end{tabular}
\end{center}
\caption{Summary of the VVV observations, sources with performed PSF photometry and selected irregular variables in each tile.}
\label{BasicInfo}
\end{table*}

Following~\cite{Medina2018} we used two main variability indices, {$ \Delta K_{\rm S}$ and $\eta$,} to select the variable star candidates. These two indices capture two fundamental properties of variable sources: the maximum change in brightness and the level of correlation among consecutive observations. The ranges of values of interest for each index were determined from their dispersion. The $\Delta K_{\rm S}$ distribution was characterized using a non-parametric fit to determine the behavior of $\Delta K_{\rm S}$ as a function of $\overline{K}_{\rm S}$. Then the dispersion, $\sigma$, in $\Delta K_{\rm S}$ as a function of $\overline{K}_{\rm S}$ was measured. Sources with amplitude above 4$\sigma$ were selected. For the $\eta$ index, we assumed that the index comes from a Gaussian distribution and used, therefore, the $\sigma$ parameter of the fitted distribution was used as a proxy for the standard deviation. Sources more than 3$\sigma$ below the mean were considered for the method, given the fact that an $\eta$ value that tends toward zero, is a strong indicator of variability~\citep{Sokolovsky2017}. 
More specifically, in this study, we selected any star with $\Delta K_{\rm S}>0.6$\,mag AND $\eta$ values $<0.95$ as a variable source.

Then, we searched for periodicity in the time series using the {\it Generalized Lomb-Scargle Periodogram} (GLS;~\citealt{Zechmeister2009}) and the {\it Information Potential Metric} $\rm Q_{m}$ (IP metric;~\citealt{Huijse2011}). We classified 177 variable stars as periodic (e.g. RRL, LPV). Their analysis will be the subject of a forthcoming paper and they are removed from the list of variable stars presented in this paper. 

\begin{figure}[h]
\begin{center}
\includegraphics[height=7cm,width=14cm]{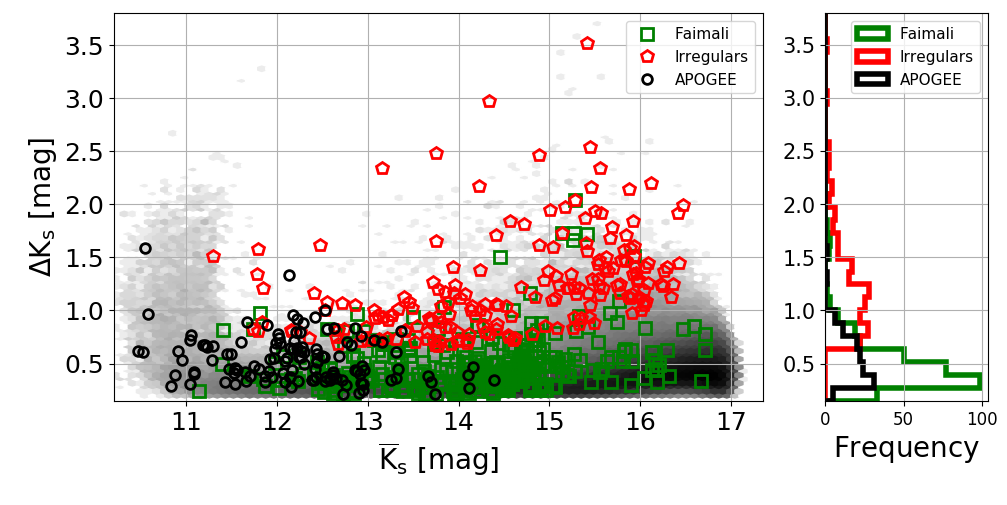}
\caption{Left panel: Amplitude $\rm \Delta K_{\rm S}$ distribution as a function of the average $K_{\rm S}$ magnitude for different samples of YSOs considered in this paper. Right panel: Amplitude histogram for the different considered datasets. Colours and shapes are displayed in the top part of the plot. In the background, the amplitudes of all objects in tile d045 are shown for comparison.
}
\end{center}
\label{Amplitude}
\end{figure}

Under the above constraints we found 196 variable sources without any periodicity. Figure~\ref{Amplitude} shows the $\Delta K_{\rm S}$ distribution for the sources found in the FoV, where red dots represent the selected {\it irregular variable sources}.

\begin{figure}[htbp]
\begin{center}
\includegraphics[height=10cm,width=17cm]{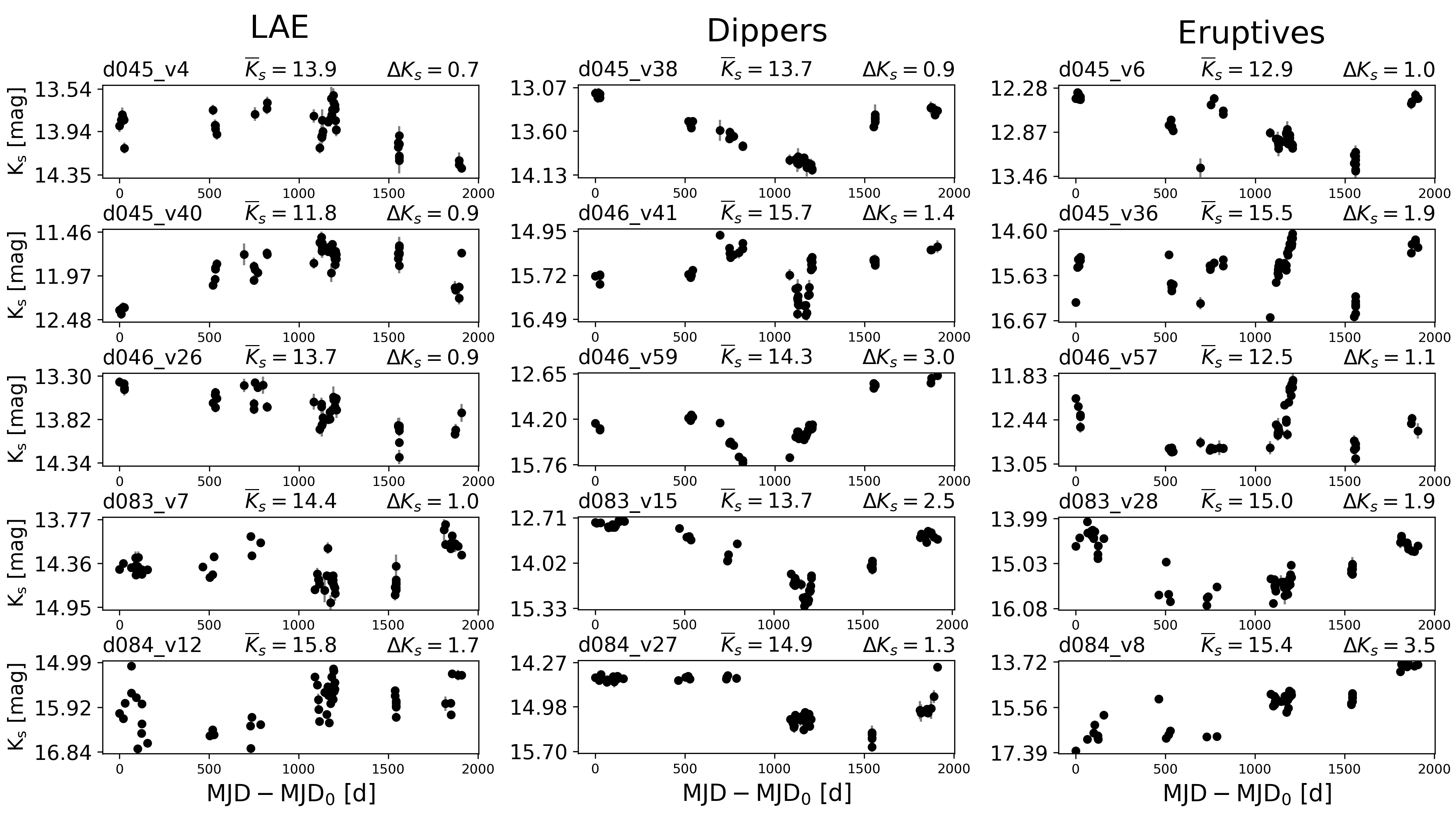}
\includegraphics[height=10cm,width=17cm]{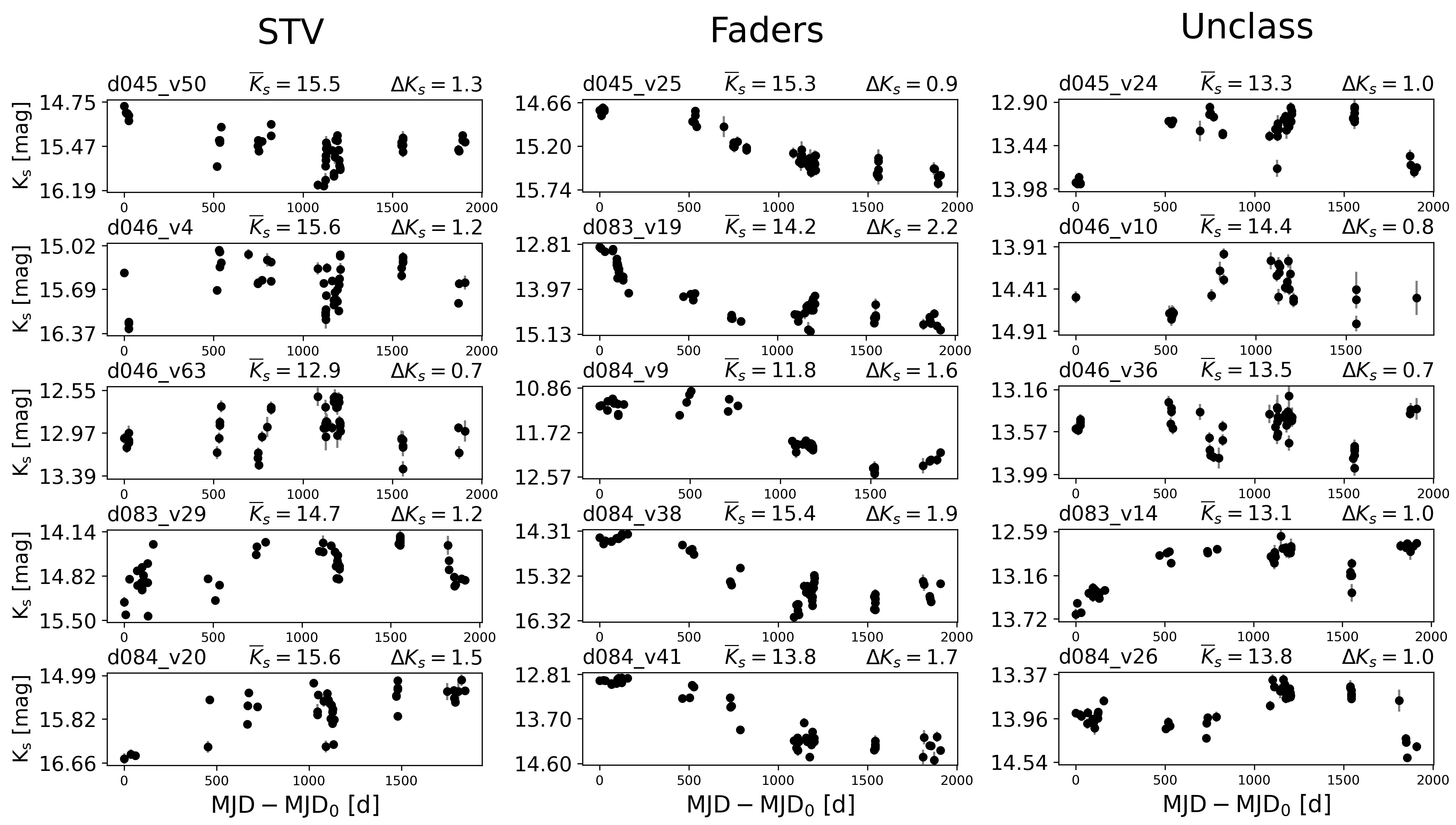}
\caption{Examples of $K_{\rm S}$-band time-series of irregular variables explained in section~\ref{GeneralProperties}. Variable class for each plot group is indicated at the top. Also, for every time-serie, the identification ID, mean magnitude $\overline{K}_s$ and amplitude $\Delta K_{\rm S}$ are shown. MJD$_0$ represent the modified julian date of the first detection for each time-series. The associated uncertainties is not visible given the amplitude of the variability.}
\label{TSs}
\end{center}
\end{figure}

Part of the complete sample of the irregular variable sources are listed in Table~\ref{v4_table}, and some examples of time-series are shown in Figure~\ref{TSs}.

\section{Comparison with previously known YSO{\tiny S} in G\,305}\label{YSOsample}

YSOs in the G\ 305 region have been investigated by ~\citet{2013PhDFaimali}, \citet{Robitaille2008},  \citet{Contreras2017a}, and \citet{Marton2019}. These data-sets were cross-matched with our time-series catalogs for d045, d046, d083 and d084 VVV tiles (using a tolerance of 0.4 arcsecs), in order to compare them with our catalog of irregular variable sources. 

\begin{description}
\item[~\citet{2013PhDFaimali}]
in this work the authors combined data from 2MASS, the first two years (2010-2011) of the VVV survey, the Spitzer Space Telescope \citep{2004ApJS..154....1W} Galactic Legacy Infrared Midplane Survey Extraordinaire \citep[GLIMPSE;][]{2003PASP..115..953B}, Multiband Infrared Photometer for Spitzer aboard the Spitzer Space Telescope \citep[MIPSGAL;][]{2009PASP..121...76C}, MSX, and the Herschel infrared Galactic Plane Survey \citep[Hi-GAL;][]{2016A&A...591A.149M} observations to conduct the first YSO census of G305. They detected 599 YSOs, but the published catalog reports only 339 of them. The authors estimated that their catalog is complete up to ${\cal M} > 2.6$ solar masses. From those, 319 have $K_{\rm S}$ time-series in our catalogs, but only six (IDs d046\_v8, d046\_v9, d046\_v22, d083\_v23, d084\_v17 and d084\_v24; see Table~\ref{v4_table}) are identified as irregular variable sources by our criteria.

\item[\citet{Contreras2017a}]
they reported 13 sources with an amplitude $\Delta K_{\rm S} > 1$ mag, that were classified either as YSO candidates or long period variables (LPVs). We recovered 12 of them. Four stars are in common with the catalog reported here (V51, V482, V43 and V480 in their list). Three sources have a well defined period and are removed from our catalog: V816 (P=616.9 d), V45 (P=584,6 d), and V49 (P=696.5 d). Another four did not fulfill the $\eta$ index criteria, having $\eta$ index larger than the average selection value: V42 ($\eta=1.4$), V44 ($\eta=1.05$), V481 ($\eta=1.25$), and v48 (with $\eta=0.93$). Source V476 has $\eta=0.6$, but an amplitude $\Delta K_{\rm S} = 0.818$ mag, thereby passing the threshold in amplitude but not passing the threshold in $\eta$.

\item[\citet{Robitaille2008}] intrinsically red sources
their catalog of extremely red objects contains sources classified as YSOs, YSOs candidates or Asymptotic Giant Branch (AGB) candidates. Those stars were selected from the GLIMPSE I and II surveys, based on their MIR color excess. There are 71 sources projected in our FoV, and nineteen of them have been identified also as irregular variable sources (see Table~\ref{v4_table}). 

\item[SIMBAD database~\citep{wenger2000}] we found 4 sources in common between this database and ours. The source d046\_v12~\citep{Borissova2016}, named DBS130 Obj1, is a member of the cluster DBS[2003]\,130. The source d046\_v40 is identified in a maser catalog of \cite{Pestalozzi2005}. The source d083\_v29 is identified as an outflow candidate in \citet{Chen2011} and \citet{Cyganowski2008}. Finally, the source d084\_v41 is a well known nova called NOVA Cen 2005 or V1047 Cen \citep{Samus2005}. It is interesting to note that the recently reported outburst in 
V1047 Cen \citep{Aydi2019} is suggested to be a dwarf nova eruption.

\item[~\cite{Marton2019}] finally, we cross-matched \cite{2013PhDFaimali}, \citet{Robitaille2008}, and our sample of irregular sources with the all-sky catalogue of YSO candidates of ~\cite{Marton2019}. The authors combined the Gaia DR2 data with Wide-field Infrared Survey Explorer (WISE) and Planck measurements using machine learning techniques. We found only 65 stars in common, which is not surprising taking into account that our sources are much fainter than the optical Gaia DR2 data, as illustrated in Figure~\ref{Kmag}. From those, 44 stars (68$\,\%$) are identified as YSOs, with $P_{Y} > 0.9$ probability. Thus, NIR studies are a useful addition to optical surveys.

\begin{figure}[ht!]
\begin{center}
\includegraphics[width=7.5cm, height=7cm]{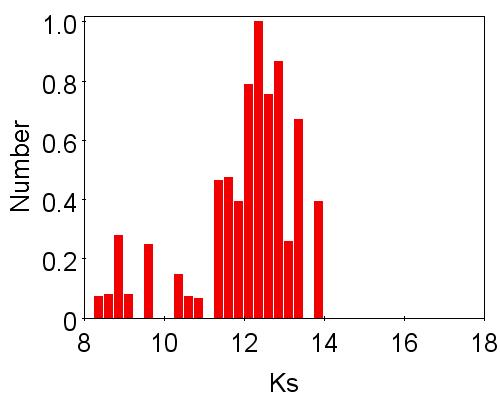}
\includegraphics[width=7.5cm, height=7cm]{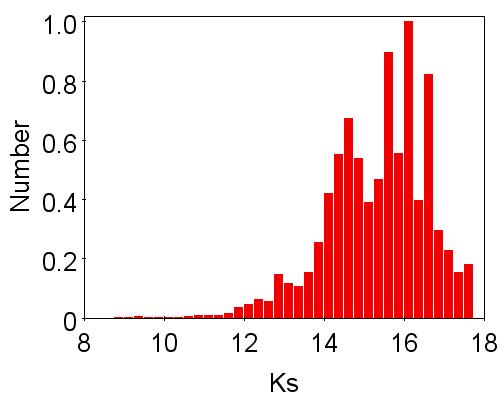}
\caption{ The histogram of $K_S$ magnitudes of ~\cite{Marton2019} catalogue of YSO candidates (left) and the combined catalog of NIR YSOs. The magnitudes are taken from the VVV database.}
\label{Kmag}
\end{center}
\end{figure}
\end{description}

The relatively small number of common objects is indicative of the very conservative and stringent selection criteria of our sample ($\Delta K_{\rm S}>0.6$\,mag and $\eta$ values $<0.95$). Nevertheless, we decided to report only highly reliable variable sources, rather than a list containing all possible candidates. We are aware that we can have omitted a true irregular variable source and thus have reduced the completeness of the sample. However, by this approach we will have few false positive detections. Thus, the total number of newly proposed irregular variable sources is 167. Most probably, according to their light curves and 
irregularity, these objects are YSO candidates. Although recent studies (e.g. \citealt {Cody2014,Contreras2017a}, Zhen et al., in preparation) found periodicity in some of the YSOs, in general the high amplitude variations do not show well-defined periods. But variability alone is not enough to confirm their status. For example, it is hard to distinguish between YSO and LPV irregular light curves. Further follow-up observations and analysis are necessary to clarify their classification. Thus, conservatively, hereafter we will continue to call these objects {\it irregular variable sources}. 

\section{Analysis and classification of the irregular variable sample}\label{GeneralProperties}

Figure~\ref{Amplitude} shows the amplitude $\Delta K_{\rm S}$ distribution of the sources as a function of the mean $\overline{K}_{\rm S}$ magnitude. In the background, as gray hexagons, the distribution of amplitudes of the whole tile d046 is shown for comparison. The amplitude interval (red open pentagons) is between $0.661 < \Delta K_{\rm S} < 3.521$ mag with an average value of $\overline{\Delta K}_{\rm S}$ = 1.22\,mag. Since the effects of saturation in VVV are substantial in the distributions for sources brighter than $\overline{K}_{\rm S} \sim 11.5$ mag and amplitudes $\Delta K_{\rm S}$ larger than 0.65 magnitudes, these sources are carefully verified. To further investigate the nature of this amplitude distribution, we compare our sample of irregular variable sources with the one from \citet{2013PhDFaimali}. Sources in this sample (green open squares) have SED classifications mainly as Class 0/I/II sources and amplitudes in the range $0.135 < \Delta K_{\rm S} < 2.04$ mag, with $\overline{\Delta K}_{\rm S} = 0.494$ mag. The amplitude distribution of APOGEE sources (black open circles) is similar, with $0.131 < \Delta K_{\rm S} < 1.58$ mag and an average value of $\overline{\Delta K}_{\rm S} = 0.54$ mag. The right panel of Figure~\ref{Amplitude} shows the histogram of the $\Delta K_{\rm S}$ distribution where our irregular sources are almost separated from the other two data sets in terms of amplitude $\Delta K_{\rm S}$ covering the higher amplitudes.

~\cite{Carpenter2001}, constructed a simple model of variations in brightness induced by cold spots, assuming a photospheric effective temperature $T_{\rm eff}\sim4000$\,K for stars with age $\sim$\,1Myr and mass $\sim$1$\cal\,M_{\odot}$. According to this model, if the spot coverage in the photosphere is assumed not to be larger than 30\,\%, the amplitude caused by this mechanism can be $ \Delta K_ {\rm S}\sim0.3 $ magnitudes. This mechanism could explain part of the amplitude distribution. On the other hand, high amplitude variability on timescales from months to decades in pre-main sequence objects is usually associated with episodic accretion events, attributed to disk instabilities typical of the early stages of young stellar objects \citep{2009ApJ...704..715V, 2019A&A...627A.154V}. 

As we discuss in section~\ref{intro}, different processes can be responsible for luminosity variations in young stars. One of the main issues in this field of research is to determine what the origin of those brightness changes is. Usually, the variability is a result of the contribution of various physical mechanisms, depending on the circumstances and environments of each particular source. In some cases, this contribution can be dominated by one of these mechanisms and then it is possible to associate the morphology of the changes as a function of time with a certain type of variability.
 
Several authors have quantified the morphology of these time series to create a connection with certain processes, but this classification depends strongly on the type of data that we are analyzing. As an example, \citet{Rebull2014}, \citet{Wolk2015}, and related papers within the YSOVAR project framework defined several types of variable stars using the MIR time-series from the Spitzer Space Telescope. The data sets from this telescope have different cadences and times for each SFR analyzed, where the periodic, quasi-periodic, and eruptive variable sources have been classified. 

For the VVV survey, NIR irregular variable sources have been split in five categories related mainly to the time-scales of the variations and the {\it overall}, most common, state: Eruptive, Faders, Dippers, Short Time-scale Variable (STV) and Low Amplitude Eruptive (LAE). These categories are summarized in Table~\ref{caracterizacion}. The classification aims to highlight the main physical process that controls the variability. The scheme was used in \citet{Contreras2017a}, \citet{Medina2018}, and references therein, in an effort to identify variable sources with VVV time-series. It is necessary to point out that, given the unevenly sampled time-series, some sources present more than one of the discussed features and will remain unclassified. Thus, for the 195 variables from our list (excluding NOVA Cen 2005, source d084\_v41, see above) we find 47 irregular variable sources with eruptive characteristics, 25 dippers, 17 faders, 24 LAE, 37 short-term variables, and 45 sources remain unclassified. Examples of these categories are shown in Figure~\ref{TSs} and the classification assigned is listed in the final table of irregular variables (Table~\ref{v4_table}). However, it is necessary to mention that some stars may have alternative classifications: d045\_v4 and d046\_v26 could be faders; d083\_v7 could be a dipper; d084\_v12 could be an STV, because short term scatter has highest amplitude; and d045\_v24 and d083\_v14 could be eruptive variables. Thus, we added the CCD analysis to the light curve morphological classification, and some spectroscopic follow-up is planned to better reveal their true category. 

\begin{table*}[htb]
\begin{center}
\begin{tabular}{p{2.3cm}p{12cm}c}
\hline\hline
\multicolumn{1}{c}{Class} & \multicolumn{1}{c}{Description} & \multicolumn{1}{c}{N$^{\circ}$ of sources} \\ \hline
Dippers & Shows one or more fading events on a timescale of months or years, followed by a return to normal brightness. & 25 \\ \hline
Eruptives  & Shows sources with outbursts with amplitude $>1$ mag and duration longer than a few days and typically at least a year. & 47 \\ \hline
Low Amplitude Eruptive  & Sources that present outbursts with amplitude lower than 1 mag and duration typically longer than a year. & 24 \\ \hline
Short Timescale Variables & Sources where the time series is dominated by fast variability on timescales from days to weeks. They also can show brief bursts in brightness on timescales of weeks. & 37 \\ \hline
Faders & The light curve is best described by a monotonic decrease in brightness on a timescale $t > 1$\,yr, sometimes following a constant brightness initially & 17 \\ \hline Unclassified & Sources without a clear trend. & 45 \\ \hline
\end{tabular}
\caption{Characterization of aperiodic VVV $K_{\rm S}$ light curves of YSOs proposed by~\cite{Contreras2017a} and extended by \citet{Medina2018}}.
\label{caracterizacion}
\end{center}
\end{table*}

\section{ $(J-H)$ vs. $(H-K_{\rm S})$ color-color diagram}\label{JHphotometry}

Additionally to the $K_{\rm S}$ photometry described above, we also obtained the $J$- and $H$-bandpass
photometry. Again, we used the $\mathtt{Dophot}$ software to perform PSF photometry in the first epoch (2010) of our VVV observations. The transformation to the standard system was performed using a linear fit between the aperture photometry from CASU and the PSF photometry. The total photometric uncertainty for each measurement is proportional to each individual PSF measurement deviation, and the standard deviation $\sigma_{\rm clip}$ obtained from the linear fit. The final uncertainties were computed as $\sigma_{\rm phot}=\sqrt{\sigma^{2}_{\rm PSF} + \sigma_{\rm clip}^{2}}$, giving $\sim0.25$\,mag at $J\sim$ 19-19.5\,mag, and $\sim0.18$\,mag at $H\sim$ 18.5-19\,mag. Due to the VVV observational strategy, both $J$- and $H$ data were taken in the same night with differences ranging from only a few minutes to not more than a few hours. The standard deviation of the linear fits and the MJD for each observation are listed in Table~\ref{JHproperties}. 
In the areas of overlap between different tiles, when both measurements were available, the average value was adopted. There are 18 sources that have been identified in two different tiles: 16 in tiles d046 and d084, one in d045-d083 and one in d083-d084. Sources having $J$ and $H$ magnitudes that were averaged across two tiles are marked in the final catalog. It should be noted that the sources do not have large discrepancies between their magnitudes in the $J$ and $H$ bands. They are in the order of 0.2 mag in $H$, and 0.1 mag in $J$ attributable to a combination of variability and measurement uncertainties. 

\begin{table*}[htb]
\begin{center}
\begin{tabular}{cccccc}
\hline\hline
Tile & $K_{\rm S}$ & $J$, $H$ & $\sigma_{\rm clip}$ & $\sigma_{\rm clip}$ & Cross-matched \\
ID  & MJD    & MJD & $J$ mag & $H$ mag &  sources   \\
\hline
d045 &    55264    &   55263    &0.062 & 0.101 &     {1,094,833} \\
d046 &    55271    &   55271    &0.047 & 0.048 &     {1,182,601} \\ 
d083 &    55254    &   55253    &0.105 & 0.062 &      {682,920}  \\ 
d084 &    55262    &   55261    &0.04 & 0.07 &      {837,668}  \\ 
\hline
\end{tabular}
\end{center}
\caption{The MJD, the standard deviation from PSF calibration $\sigma_{\ rm clip}$ and the cross-matched sources for all considered tiles in G\,305.}
\label{JHproperties}
\end{table*}

As pointed out by ~\citet{hindson2012}, the G\,305 region contains dust clumps spread over the field, which produce extinction towards the stellar sources. A widely used technique to characterize such highly reddened sources is their position in the CCD. Particularly, as disused in \cite{Lada1992} on the NIR ($J-H$, $H-K_{\rm S}$) plane, the stars with intrinsic excess emission, heavily reddened sources, and stars with normal photospheric colors can be distinguished from each other. For further analysis, we use ($J-H$, $H-K_{\rm S}$) CCDs for the 2010 epoch. To create the CCD we used the closest $K_{\rm S}$ epoch in time to the $J$ and $H$ images. For the tiles d045, d083 and d084 the images were observed with one day of separation, meanwhile for d046 $K_{\rm S}$ data were taken on the same night as the $J$ and $H$ photometry. To perform the cross-match between catalogs, we used a tolerance of 0.4 arcseconds in all the selected regions. Table~\ref{JHproperties} contains the MJD for each image and the quantity of cross-matched sources for each VVV tile. 
Then, we divide the ($J-H$, $H-K_{\rm S}$) plane following the method of ~\cite{Sugitani2002} and \cite{Ojha2004}. The sources are analyzed with respect to the loci followed by main-sequence stars, giant stars, and classical T-Tauri stars defined in~\cite{Bessell1988} and~\cite{Meyer1997}. Three regions can be distinguished as follows: 

\begin{enumerate}
\item F-region: We define the region between the reddening vectors of main-sequence stars and giant stars as the F-region. These objects are considered to have at most a small NIR excess, likely corresponding to Class III or Class II YSOs. Some of these could be foreground or background sources. 

\item T-region: The T-region is defined to be between the projected reddening vector from the end of the T-Tauri locus~\citep{Meyer1997} and the main-sequence dwarf reddening vector. Sources located here are considered to have a substantial NIR excess and are likely to be Class II objects, also known as T-Tauri stars. Some Herbig Ae/Be sources with small NIR excess are also considered to lie in the T-region.

\item P-region: The region to the right of the reddening vector extending from the end of the T-Tauri locus is called the P-region. The projected sources in this region are more likely Class I objects, sources with thermal emission from the circumstellar envelope, and protostar-like and/or Herbig Ae/Be objects.
\end{enumerate}

\begin{figure}[htbp]
\begin{center}
\includegraphics[height=8cm,width=15cm]{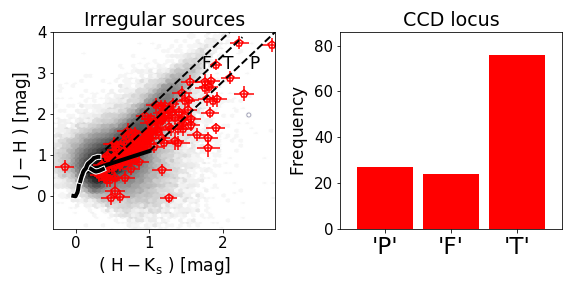}
\\
\includegraphics[height=8cm,width=15cm]{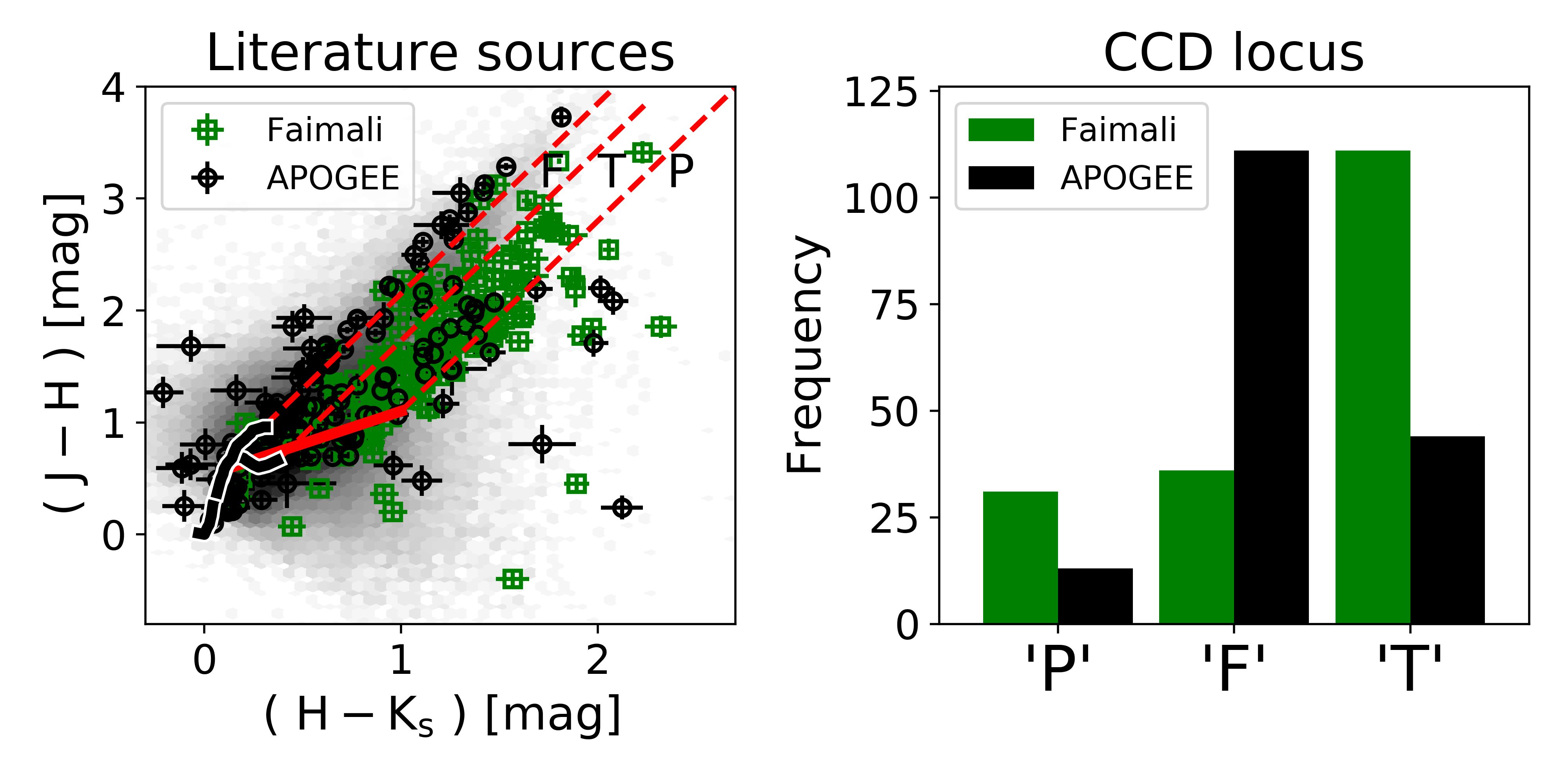}
\caption{Left panels: CCDs for all selected tiles in VVV, displaying the so-called F, T and P-regions. In the background of the CCDs, as blue hexagons, the color distribution of tile d046 is shown as a comparison. Right panels: The histograms with the CCD classification. 
}
\label{CCD_irregulars}
\end{center}
\end{figure}

In general, only a subset of all objects has been detected in $J$ and $H$ bands, but still 130, 179 and 127 of our irregular variables, the \cite{2013PhDFaimali} sample, and the APOGEE sample, respectively, have reliable $JH$ measurements and can be placed in the CCD (see Fig.~\ref{CCD_irregulars}). Sources have been displayed in a CCD for each selected VVV tile, where the entire population of each tile is plotted in the background. Solid black lines correspond to unreddened main-sequence stars and giant stars sequence \citep{Bessell1988}. Solid red line represents the intrinsic T-Tauri locus \citep{Meyer1997}. The reddening vectors using the~\cite{RiekeLebofsky1985} extinction law and assuming a visual extinction $A_{V}=15$ mag for each evolutionary track are also shown. 

In total, we were able to classify 116 of our irregular variable sources. 

Only a small fraction of them falls inside the F-region between the reddening vectors projected from the intrinsic color of main-sequence stars and giants. The majority are near to the red border of this region and probably are Class III/Class II sources with small NIR excess. The rest of the sample falls redward of the F-region, thus indicating the presence of intrinsic infrared excess emission, a distinctive feature of young sources with circumstellar matter.

T-region sources are located redward of the F-region, but blueward of the reddening line projected from the red end of the T Tauri locus. They are considered to be mostly classical T-Tauri stars (Class II objects) with clear NIR excess. This is consistent with the type of stellar objects that we expect to identify given their IR variable nature. To support this interpretation, we built the membership histogram for all selected sources (top histogram in figure~\ref{CCD_irregulars}). About 
60\,\% of identified irregular sources are located inside the T-region. 

Protostellar sources (i.e., Class I YSOs) are expected to lie redward of the reddening vector of the classical T Tauri (CTT) locus, in the so-called P-region, given they are intrinsically red due to their evolutionary state. Some $\sim$20\,\% of the whole sample falls in this region. 

A substantial part of the \citet{2013PhDFaimali} YSO sample ($\sim$63\,\%) lies in the T-region. The right panel of Figure~\ref{CCD_irregulars} shows the membership histogram of each category.

Most stars from the APOGEE sample fall in the F-region. Also, in the CCD they populate areas characterized by low values of visual extinction, close to unreddened loci of main sequence stars and giants. 

The above membership assignments of the sources will only be considered as a preliminary estimate of their evolutionary stage.

\section{MIR photometry and SED class}
In order to better analyze the irregular variables, we cross-identified our sample with the GLIMPSE IRAC catalog. IRAC provides the highest spatial resolution of any mid-IR imager for panoramic surveys over wavelengths from 3 to 9 $\mu$m. The IRAC photometry is much more reliable near the Galactic plane than the WISE photometry, which was severely limited there by detector saturation and especially by source confusion. We also made a cross-identification with 24 $\mu$m Spitzer/MIPS band, which unfortunately is not available for the vast majority of the point sources detected in GLIMPSE. For this reason, we made an additional cross-match with the WISE catalog, from which we used only W4 photometry. 

In order to assess the evolutionary stages of YSOs in our sample, we used the  infrared spectral index, $\alpha$. The value of the index depends on the spectral range. The wider the range towards the MIR, the higher
the sensitivity to different kinds of disks. Additionally, the calculation of the spectral index is also affected by the reddening \citep{2005ApJ...619..931I}. To estimate values that are minimally affected by the reddening, we use the wavelength range from 4.5 $\mu$m to 24 $\mu$m. The interstellar extinction in these bands is smaller than at shorter wavelengths, and the reddening curve is flatter. We obtained $\alpha$ following the procedure and using the equations given in \citet{2020arXiv201112961K}:

$$\alpha_{[4.5]-[24]} \approx 0.55([4.5]-[24])- 2.94,$$

$$\alpha_{[4.5]-W4} \approx 0.58([4.5]-W4)-2.92,$$

$$\alpha_{[4.5]-[8.0]} \approx 1.64([4.5]-[8])-2.82.$$
All data from GLIMPSE, MIPS and WISE are given in Table\,\ref{v4_table}.

As in \citet{2020arXiv201112961K}, we prefer the $\alpha$ estimate based on [4.5] -- [24], followed by [4.5] -- W4, and finally [4.5] -- [8.0]. 
The morphology of the ([4.5] -- [5.8] vs. [5.8] -- [8.0]) CCD shows that in our sample, there are no YSOs presenting strong silicate absorption or PAH emission. 

From the entire list of 196 variables, there are IRAC, MIPS, and/or WISE data for 116 objects ($\sim$ 59\% of all), allowing us to obtain the index $\alpha$ for all of them. The $\alpha$ values and the corresponding Class are given in Table\,\ref{v4_table}. 

\begin{figure*}[ht!]
\begin{center}
\includegraphics[height=6cm,width=17cm]{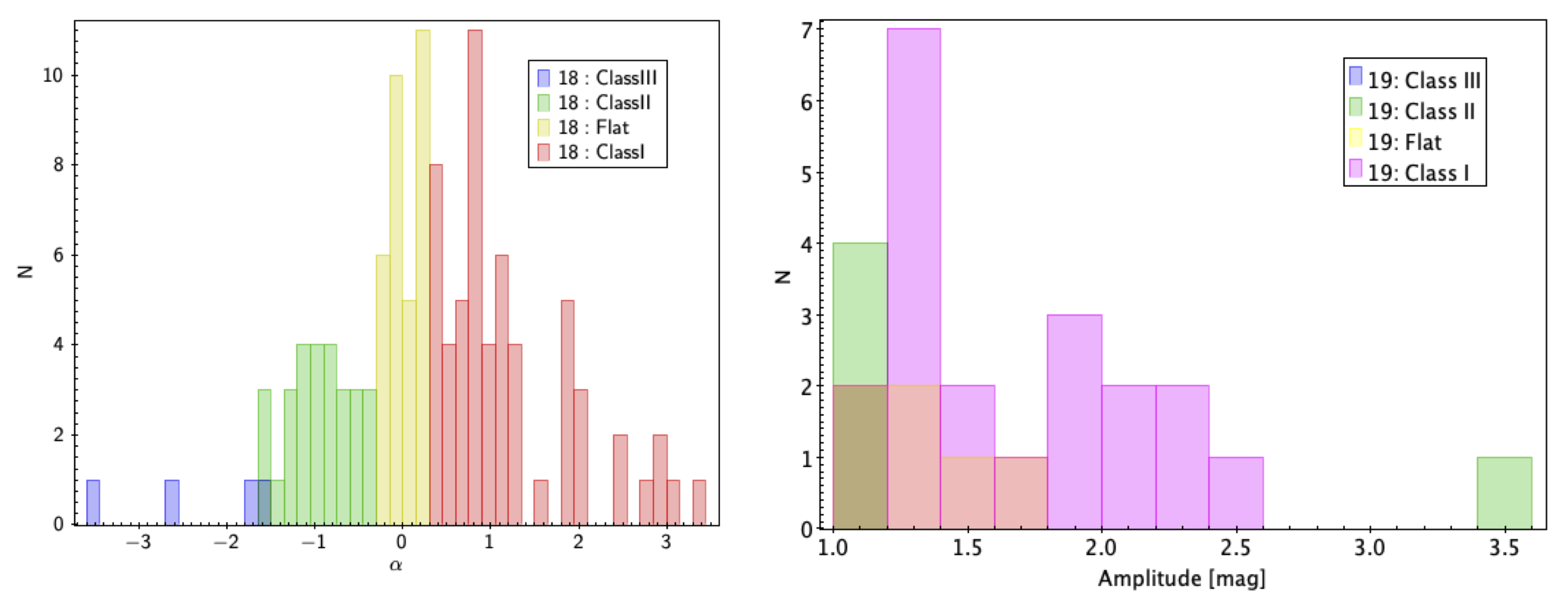}
\caption{Left panel. Histogram distribution of spectral indices for 116 YSOs with MIR photometry, subdivided into YSO class using the customary demarcations at $\alpha = -1.6, -0.3$, and $0.3$. Right panel. Histogram distribution of the amplitudes of the objects classified as {\it Eruptive}. }
\label{alpha_hist}
\end{center}
\end{figure*}

Figure\,\ref{alpha_hist} (Left panel) shows the distribution of the spectral indices for YSOs. Subdivision into YSO classes is made using the limits at $\alpha = -1.6, -0.3$, and $0.3$ \citep{1994ApJ...434..614G}. Based on these criteria, there are 58 Class I ($\alpha > 0.3$), 32 Flat spectrum $(0.3 \geq \alpha > -0.3)$, 28 Class II $(-0.3 \geq \alpha > -1.6)$, and only 4 Class III $(\alpha \geq -1.6)$ YSOs. 

The shape of the distribution depends on the relative number of the YSO classes. In general, Class II/III YSOs should be more common than Class I/Flat SED YSOs due to the longer lifetimes of the later evolutionary stages \citep{2009ApJS..181..321E}. But our study, based on variability, is initially limited by the amplitude ($\Delta K_{\rm S}\ge0.6$), which is expected to exclude higher proportions of Class III and Class II YSOs than Class I YSOs. This is clearly visible in Figure\,\ref{alpha_hist} (Left Panel). So, we are much more sensitive to the younger YSO population. 

Additionally, we analyze the amplitude distribution for the YSOs classified as {\it Eruptive}, shown in Figure\,\ref{alpha_hist} (Right panel). Not surprisingly, Class III objects are missing in this figure. The Class II YSOs show lower amplitudes of about 1\,mag. The Flat spectrum objects are concentrated between 1 and 1.6\,mag, and the Class I, as expected, span the whole range from 1 to 2.5 mag. The only exception in this classification is d084-v8, which is classified as Class II, but shows the highest amplitude of 3.52\,mag. Most probably, it is a result of a wrong $\alpha$ index, obtained only by the WISE W4 magnitude and missing [8.0] and [24] measurements. Also, the NIR classification for this object is missing, because it is too faint and too red, and has no $J$ and $H$ measurements.

\section{APOGEE-2 sample}

APOGEE-2 \citep{Majewski2017} is a second generation multi-object NIR spectrograph \citep{Wilson2019}, mounted on the 2.5 meter du Pont Telescope at Las Campanas Observatory, Chile \citep{gunn2006}. It is the successor to APOGEE and it is part of the Sloan Digital Sky Survey IV \citep{Blanton2017}. The observed spectral range is 1.51-1.70 $\mu$m, with resolution R=22500. Approximately 300 stars can be observed per plate, typically assigning about 250 fibers for science targets, 35 for sky, and 15 for telluric stars, with the fiber collision limit of about 70$''$. To design the strategy for spectroscopic follow-up, we assigned priorities for observation based on the preliminary classification of the objects: 59\,\% are stars with infrared variability in $K_{\rm S}$ band from our search (priority 1); 21\,\%, (priority 2) are from the \citet{2013PhDFaimali} catalog; 8\,\% (priority 3) are extremely red sources from \citet{Robitaille2008}; 3\,\% show X-ray emission \citep[Chandra Source Catalog][]{Evans2010}; 7\,\% are probable members of the clusters VVV CL\,021, CL\,022, CL\,023 and CL\,024 \citep{Borissova2011}, and three stars (2\,\%) are classified as high proper motion candidates \citep{Kurtev2017}. The last three groups were added as ``fillers'', during the target selection process. The APOGEE-2 observations of G\,305 were carried out on April 14, 2017 (under external CNTAC program No. CN2016B90), with 1 hour exposure time.
The data were processed by the APOGEE Stellar Parameter and Chemical Abundances Pipeline \citep{Nidever2015, Garciaperez2016}, which includes basic data reduction, combination of spectra from multiple visits, and measurement of radial velocities, stellar parameters and elemental abundances.\footnote{For more details see http://www.sdss.org/dr14/algorithms/} 

In total, 127 stars were observed with APOGEE-2. Their spectra are analyzed using the TONALLI (L. Adame, in prep.) spectral classification code. In brief, the TONALLI code is based on the {\it Asexual Genetic Algorithm} \citep{2009A&A...501.1259C} to find the closest match among a synthetic model grid using a multi-dimensional approach based on a set of parameters. The code uses the Phoenix~\citep{Allard2013} atmosphere model grid with "solar" abundances between $-0.1<{\rm M/H}<0.6$ and $-0.1<\alpha/{\rm Fe}<0.1$, after pre-selecting ``low temperature'' star candidates with $T_{\rm eff}<8000$~K. As a result of the best fit, it derives the abundances [Z], [$\alpha$/Fe], $\log g$, $v\sin i$, and radial velocities (RV). 

\begin{figure*}[ht!]
\begin{center}
\includegraphics[height=6cm,width=17cm]{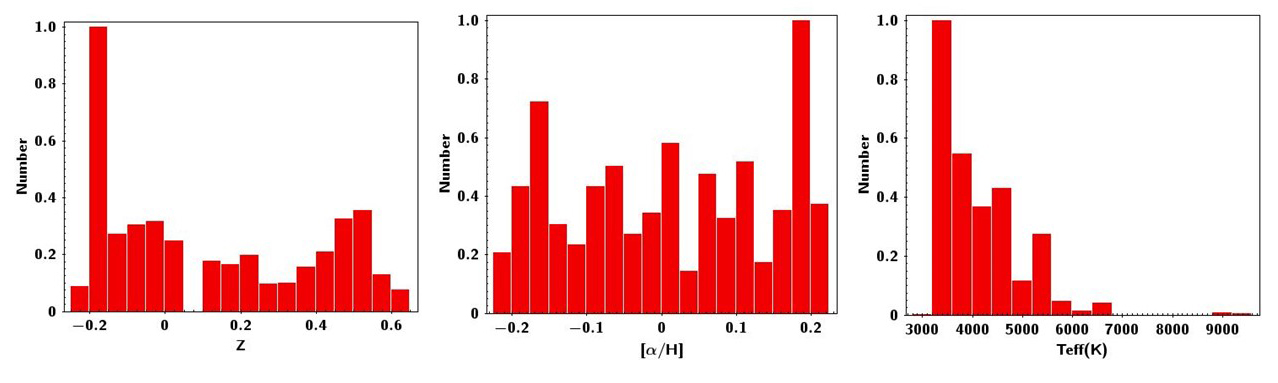}
\caption{Distribution of YSOs according to their fundamental parameters derived with APOGEE-2: Histograms of metallicity ($Z/Z_{\odot}$), $\alpha$ elements, and $T_{\rm eff}$. All parameters are measured by the TONALLI tool.}
\label{apogee_parameters}
\end{center}
\end{figure*}

In the case of G\,305, 122 stars are classified by the tool as YSOs, but 25\,\% of them have S/N between 2 and 10 and lower weight (between 0.2 and 0.5) was assigned for their parameters during the analysis. 
Figure~\ref{apogee_parameters} shows the distribution of their fundamental parameters. The measured effective temperature $T_{\rm eff}$ peak is around 4000K, in agreement with ~\cite{Carpenter2001}. There is a peak in $[\alpha/{\rm Fe}]$ around 0.2 but the mean value of $[\alpha/{\rm Fe}]$=0.02$\pm0.13$. The $Z/Z_{\odot}$ distribution has a peak around $-$0.2, but a second peak around 0.5 can be also identified. The mean $Z/Z_{\odot}$ of 0.11 $\pm0.27$ reveals that the YSOs in G\,305 are slightly super-solar in metal abundance. The radial velocity peak is around $-$41 km/s.

A cross match with the Gaia DR2 proper motion, parallax, and distance \citep{Bailer-Jones} catalogs found 104 stars in common. The average distance is calculated as $3.7\pm 0.7$\,kpc. Comparison with radial velocities and distance of the OB stars in the region \citep{Borissova2019} shows no significant differences from a kinematic point of view between the two stellar populations.

With regard to the variability of these YSOs, Figure~\ref{Amplitude} shows the distribution of their amplitudes. As can be seen, in general, they have smaller amplitudes compared to the irregular variable sources from our list (marked with red symbols). Nevertheless, all of them have amplitudes between $0.209 < \Delta K_{\rm S} < 1.587$\,mag -- an average of 0.54\,mag --, thus showing some kind of variability. We could not find any correlation between fundamental parameters and the amplitude of variability. We calculated the Pearson correlation coefficients of [M/H] vs. Amplitude to be $\approx -0.4$; [Fe/H] vs. Amplitude to be $\approx -0.4$; $[\alpha/{\rm H}]$ vs. Amplitude to be $\approx -0.4$; T$_{\rm eff}$ vs. Amplitude to be $\approx -0.4$; and $\log g$ vs. Amplitude to be $\approx -0.4$. Taking into account the relatively small number of stars, we found these correlation coefficients to be statistically insignificant.

\begin{figure}[!tbp]
\begin{center}
\includegraphics[width=10cm,height=6cm]{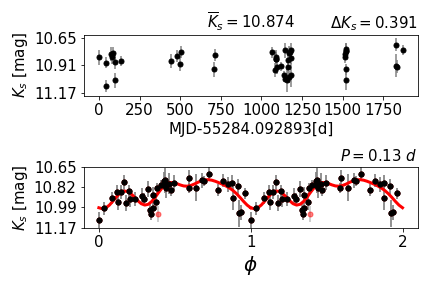}
\caption{Time-series (top panel) and folded light curve with a P=0.13 [days] (low panel) of source 2M13121845-6237309. The period has been found using the IP metric~\citep{Huijse2011}. The red line in the folded light curve represents the harmonic fit, and red points are the rejected measurements of the 3$\sigma$ clipping of the fit.}
\end{center}
\label{binary}
\end{figure}

As has been pointed out, three stars (VVVJ\,13081183-6317548; VVVJ\,13081329-6242491 and 2M13182865-6253076) classified as high proper motion objects ~\citep{Kurtev2017} were observed by APOGEE as "fillers". Although far from the main topic of the paper, we analyzed them and will briefly mention here.  
The Gaia DR2 data shows that they can be measured in the optical waveband too and indeed they have large proper motions (Table~\ref{high_pm}). Comparing the Gaia temperature vs. the~\cite{Straizys1981} spectral type and the Spectroscopic type vs. $\log g$ (measured by APOGEE spectra) calibrations, we derived their spectral types. Although this classification has an uncertainty of 2-3 subtypes, these are identified as K and early M-type stars rather than ultracool dwarfs. From the X-ray emitters, only 2M13121845-6237309 has a tentative period of 0.13 days and its light curve resembles that of a binary star (Figure~\ref{binary}).

\section{Spatial distribution around the G\,305 SFR}\label{SpatialDistribution}

\begin{figure}[htbp]
\begin{center}
\includegraphics[height=4.5cm,width=\linewidth]{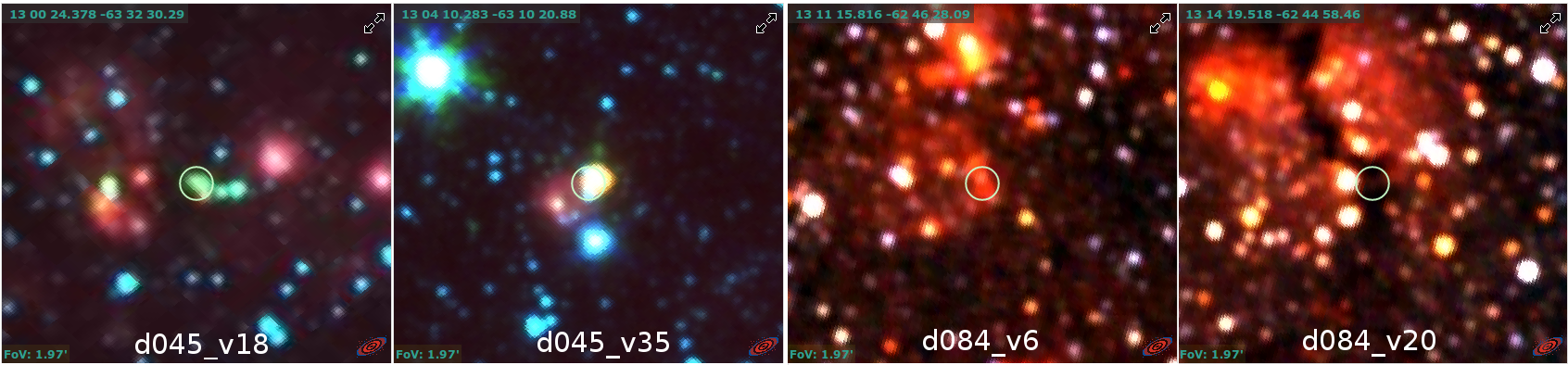}
\caption{Examples of sources projected in a different environment overplotted on IRAC (two on the left) and 2MASS (two on the right) bands false color images.}
\label{VariableEnviroment}
\end{center}
\end{figure}

The location of the identified irregular variable sources can be seen on the panoramic view of the G\,305 star-forming complex shown in Figure~\ref{FoV}. The main data sets considered in this study are marked with different symbols and colors. The false-color view provided by the Spitzer Space Telescope InfraRed Array Camera (IRAC) at mid-infared wavelengths of 3.6$\mu$m to 8$\mu$m is in the background. 
As can be seen, the selected irregular sources extend throughout the whole area. However, the density distribution is different: the sources in tiles d046 and d084 are mostly projected on the red filaments across the image. These trace the polycyclic aromatic hydrocarbons (PAH) emission, excited by surrounding interstellar radiation and become luminescent at wavelengths near 8.0$\mu$m. The PAH emission is typically found from photo-dissociated regions \citep[e.g.][]{1997ARA&A..35..179H}. 
On the other hand, the sources in the tiles d045 and d083 show a nearly homogeneous distribution. Thus, the distribution of the young sources in G\,305 clearly trace the ongoing star-formation zones. Some individual examples are shown in Figure~\ref{VariableEnviroment}, where the sources d045\_v18 and d045\_v35 are projected in the IRAC bands on green nebulosity, indicating an outburst or jet that is bright at 4.5~$\mu$m. The right panels show the d084\_v6 and d084\_v20 sources projected on nebulous regions in 2\,MASS images. 

\begin{figure}
\begin{center}
\includegraphics[height=13cm ,width=6.8cm]{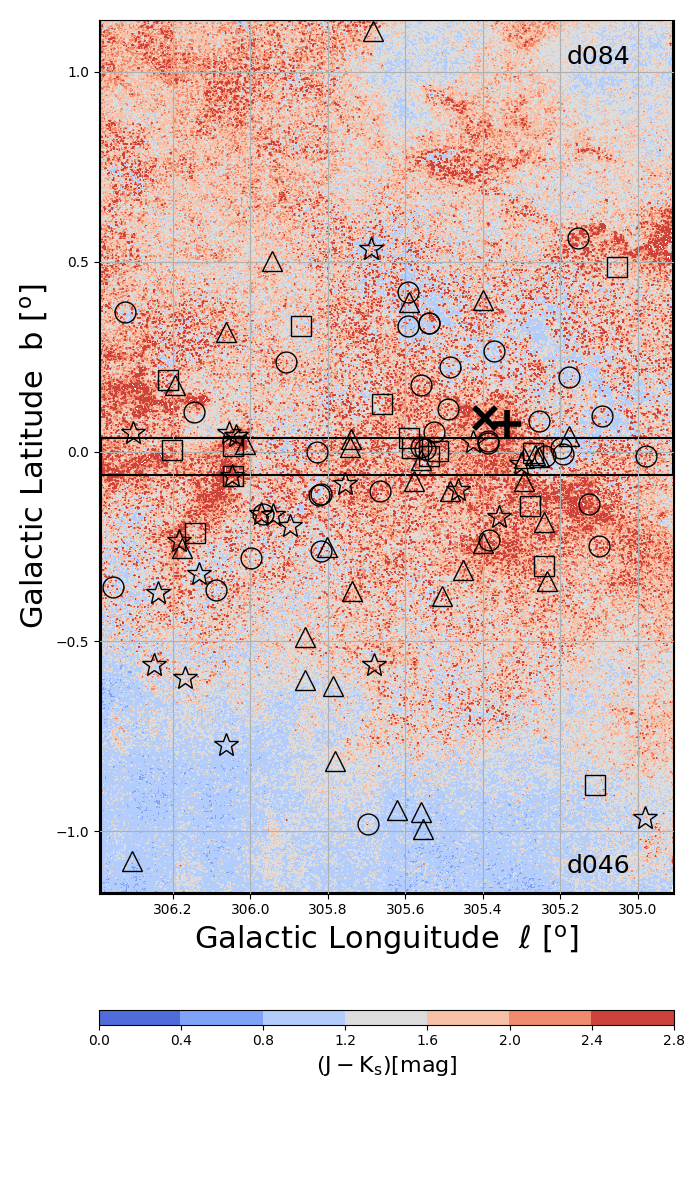}
\includegraphics[height=13cm, width=8cm]{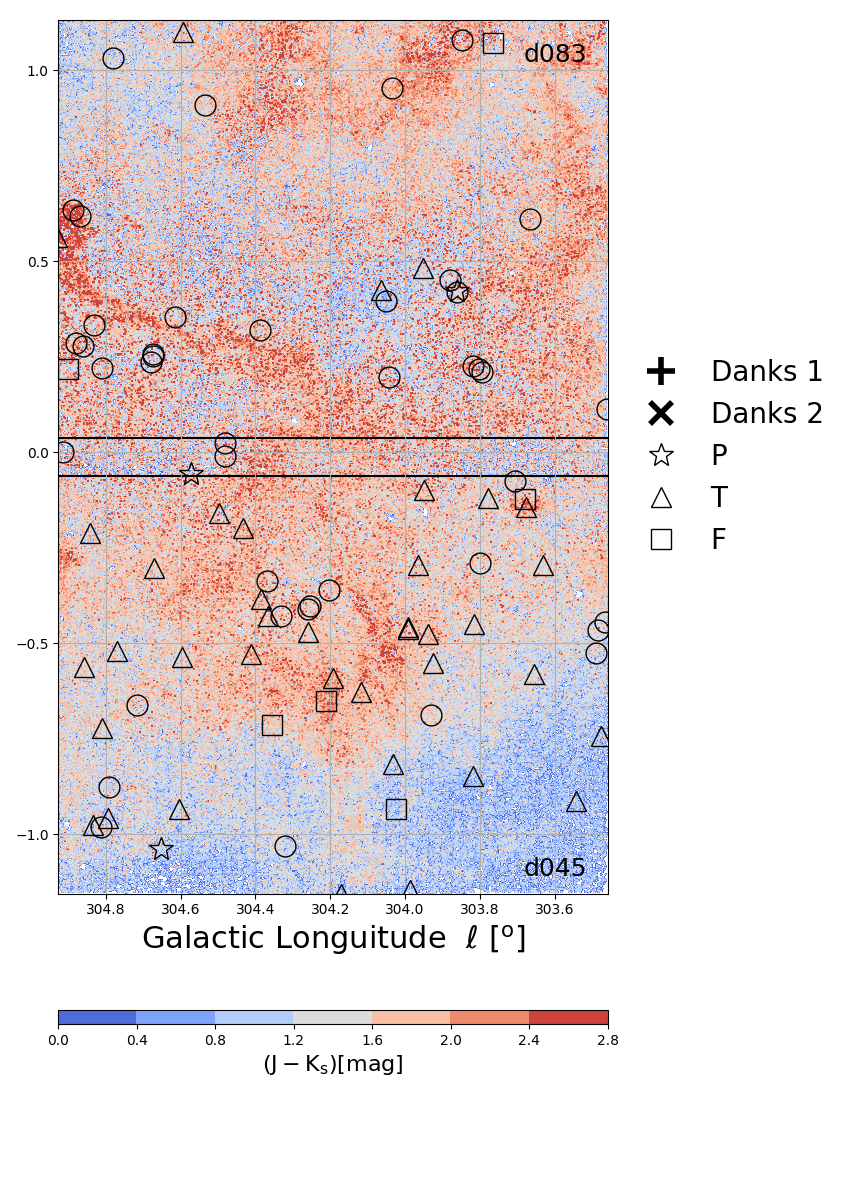}
\caption{Distribution of the observed $J-K_{\rm S}$ colors of the stars in G\,305 region with variable irregular sources overplotted. Open circles, represent the sources without color. The horizontal lines mark the overlapping zone of the tiles d084-d046, and d083-d045 respectively. The symbols for the P, T and F-objects, as well as for the position of the clusters Dansk\,1 and 2, are given on the right.}
\label{jk}
\end{center}
\end{figure}

The spatial projection of the classified variables can be also illustrated with the smoothed surface $(J-K_{\rm S})$ color distribution shown in Figure~\ref{jk}. Considering that class I/II objects (comparing with class III) are more likely to show variability due to their instability while evolving towards the main-sequence~\citep{Rice2015, Contreras2017a} one should expect that they are distributed and concentrated in dust and gas over-densities. This trend is clearly visible in G\,305 region.  
 
\begin{figure}[ht!]
\begin{center}
\includegraphics[width=15cm, height=12cm]{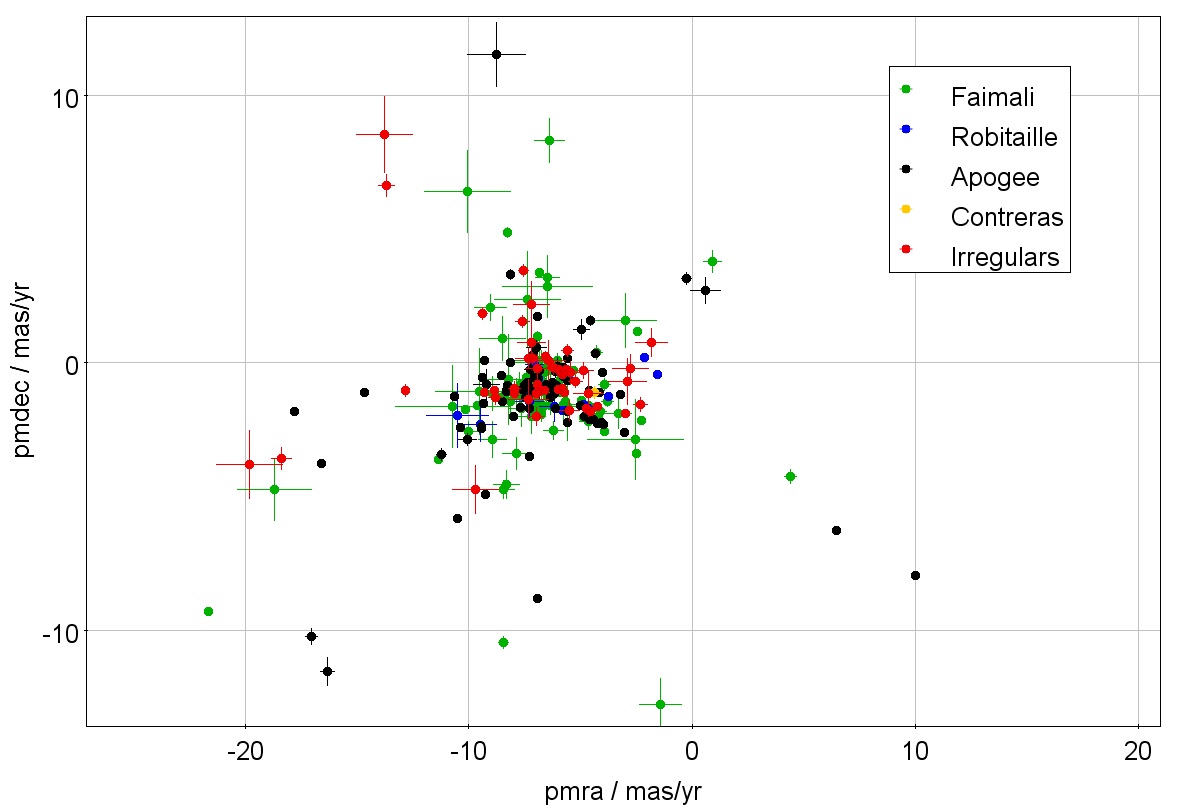}
\includegraphics[width=7.5cm, height=7cm]{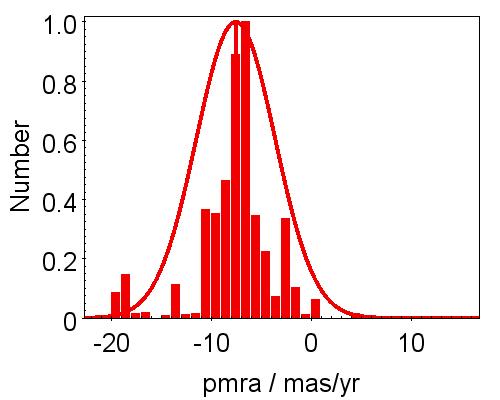}
\includegraphics[width=7.5cm, height=7cm]{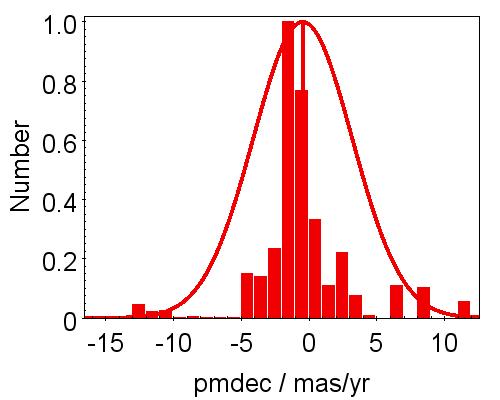}
\caption{Up: The kinematic distribution of YSOs with measured proper motions from Gaia DR2. The different data sets are
shown in colors, explained in the legend (see text). Down: The histograms of $\mu_\alpha\cos\delta$ and $\mu_\delta$ with Gaussian function overplotted. }
\label{all_pm_gaia}
\end{center}
\end{figure}

As mentioned before, in this paper we analyzed several catalogs, which contain YSOs or YSO candidates: \citet{2013PhDFaimali}, \citet{Contreras2017a}, \citet{Robitaille2008}, the APOGEE-2 confirmed YSOs, and the irregular variable sources derived from the VVV $K_{S}$ band analysis, comprising a total of 700 stars. We will consider this catalog as a statistically representative sample of YSOs and candidates in the region. The sample was cross-identified with Gaia DR2, yielding 295 stars with proper motion measurements. As can be seen in Figure~\ref{all_pm_gaia}, their distribution is typical for a structure belonging to the galactic disk. The stars form a compact group centered on $\mu_\alpha\cos\delta \approx-7.5\,{\rm mas\,yr
^{-1}}$ and $\mu_\delta \approx -0.5\,{\rm mas\,yr
^{-1}}$. 

Moreover, the distribution projected on the sky is not random, but in fact most of the sources are concentrated within the central cavity formed by two massive clusters, namely Danks\,1 and 2 (\citealt{2013PhDFaimali}; see also Figure~\ref{FoV}, where different catalogs are color-coded). The mean surface density of the YSOs in the region is calculated to be 0.025 YSOs/pc$^{-2}$, while the central cavity contains 10 times more objects per area (0.238 YSOs/pc$^{-2}$), clearly tracing the propagation of star formation in the region.

\section{Discussion and Summary.}

G\,305 is a relatively close-by, luminous giant HII region with active star formation. Using the time-series photometry in the $K_S$ bandpass provided by the VVV survey for the sources in this region, we generated and analyzed 3,570,064 unevenly sampled light curves over a 5-year period. The variability indices $\Delta K_{\rm S}>0.6$\,mag and $\eta$ values $<0.95$ are used in order to select variable star candidates. We found 373 high-amplitude variable candidates in the region: 177 sources show periodicity in their variability, while 196 seem to show irregular variations in their brightness. After cross correlation with catalogs of YSOs available in the literature and SIMBAD, 167 stars are identified as new irregular variable stars. Inspection of the light curves indicates that these objects are good YSO candidates. On the basis of the morphology of their light curves, the sources are classified using~\cite{Medina2018} scheme: Eruptive, Faders, Dippers, STVs, and LAEs. It is interesting to mention, that from classified variables the biggest group  (31\%) contains Eruptive sources, confirming that the eruption is a frequently physical mechanisms for the brightness changes.

Additionally, the YSOs from the different catalogues from the literature  ~\cite{Robitaille2008,2013PhDFaimali,Contreras2017b} are analyzed. The $(J-H)$ vs. $(H-K_{\rm S})$ CCDs are constructed and the position of variable sources on the diagrams are connected with their evolutionary status (Class I, Class I/II and Class III). About 60\,\% of the identified irregular sources are located at the so-called T-region. This region encompasses variables with large NIR excess, and is usually regarded as the probable location of Class II systems. In order to improve this classification, we cross-identified our sample of 196 non-periodic variables with GLIMPSE, MIPS, and WISE catalogs. There are available data for 116 of them. Using the infrared $\alpha$ index, we classified these 116 objects as follows: 58 Class I, 32 flat spectrum, 28 Class II, and only 4 Class III YSOs. Thus, the majority of highly variable YSOs with mid-IR detections are classified as Class I systems, even if they are located in the T- region of the near IR two colour diagram.  (Sources lacking mid-IR detections are of course more likely to be Class II or Class III systems). The high proportion of Class I YSOs clearly reflects the ongoing star formation in G\,305 region.
Comparing with the OB stellar population, we can support the \citet{Clark2004} suggestion of  sequential star formation.  

In total, 127 stars were observed with the APOGEE-2 multi-object NIR spectrograph. From them, 122 stars are classified by the TONALLI code as YSOs. The mean $Z/Z_{\odot}$ of 0.11 $\pm0.27$ suggests that the YSOs in G\,305 are slightly super-solar in metal abundance. The radial velocity peak is around $-$41 km/s. The average distance of $3.7\pm 0.7$\,kpc is calculated, on the base of Gaia DR2 distance catalog. Thus, we do not find any significant differences from a kinematic point of view between the YSOs and OB stellar population ~\citep{Borissova2019} in the region. All stars with APOGEE-2 spectra are variable, with amplitudes between $0.209 < \Delta K_{\rm S} < 1.587$\,mag in the $K_{\rm S}$ band. 

The spatial distributions of the combined catalog of 700 YSOs in G\,305 shows that the distribution is not random, but in fact the YSOs are concentrated within the central cavity formed by the massive clusters Danks\ 1 and 2. We find the mean surface density of YSOs to be 0.025 YSOs/pc$^{-2}$, while the central cavity contains 10 times more objects per area (0.238 YSOs/pc$^{-2}$).

\section{Acknowledgments}
We thank the referee for carefully reviewing the manuscript and for the valuable suggestions and comments. We gratefully acknowledge data from the ESO Public Survey program ID 179.B-2002 taken with the VISTA telescope, and products from the Cambridge Astronomical Survey Unit (CASU), also on data products from observations made with ESO Telescopes at the La Silla, Paranal Observatory under program ID 177.D-3023, as part of the VST Photometric H$_\alpha$ Survey of the Southern Galactic Plane and Bulge (VPHAS+). 
This work was funded by ANID – Millennium Science Initiative Program – ICN12\_009 awarded to the Millennium Institute of Astrophysics (MAS). N.M. acknowledges support through a Fellowship for National PhD students from ANID, grant number 21181952. J.A.-G. acknowledges support from Fondecyt Regular 1201490. A.B. acknowledges funding by ANID, Millennium Science Initiative, via the N\'ucleo Milenio de Formaci\'on Planetaria (NPF) and from FONDECYT Regular 1190748. A. Roman-Lopes acknowledges financial support provided in Chile by Comisi\'on Nacional de Investigaci\'on Cient\'ifica y Tecnol\'ogica (CONICYT) through the FONDECYT project 1170476 and by the QUIMAL project 130001. C. Rom\'an-Z\'u\~niga acknowledges support from project CONACYT CB2018 A1-S-9754 and project UNAM-PAPIIT IN112620, Mexico.

Funding for the Sloan Digital Sky Survey IV has been provided by the Alfred P. Sloan Foundation, the U.S. Department of Energy Office of Science, and the Participating Institutions. SDSS-IV acknowledges
support and resources from the Center for High-Performance Computing at
the University of Utah. The SDSS web site is www.sdss.org.

SDSS-IV is managed by the Astrophysical Research Consortium for the 
Participating Institutions of the SDSS Collaboration including the 
Brazilian Participation Group, the Carnegie Institution for Science, 
Carnegie Mellon University, the Chilean Participation Group, the French Participation Group, Harvard-Smithsonian Center for Astrophysics, 
Instituto de Astrof\'isica de Canarias, The Johns Hopkins University, Kavli Institute for the Physics and Mathematics of the Universe (IPMU) / 
University of Tokyo, the Korean Participation Group, Lawrence Berkeley National Laboratory, 
Leibniz Institut f\"ur Astrophysik Potsdam (AIP), 
Max-Planck-Institut f\"ur Astronomie (MPIA Heidelberg), 
Max-Planck-Institut f\"ur Astrophysik (MPA Garching), 
Max-Planck-Institut f\"ur Extraterrestrische Physik (MPE), 
National Astronomical Observatories of China, New Mexico State University, 
New York University, University of Notre Dame, 
Observat\'ario Nacional / MCTI, The Ohio State University, 
Pennsylvania State University, Shanghai Astronomical Observatory, 
United Kingdom Participation Group,
Universidad Nacional Aut\'onoma de M\'exico, University of Arizona, 
University of Colorado Boulder, University of Oxford, University of Portsmouth, 
University of Utah, University of Virginia, University of Washington, University of Wisconsin, 
Vanderbilt University, and Yale University.

\software{$\mathtt{Matplotlib}$~\citep{Hunter2007}, $\mathtt{NumPy}$~\citep{Oliphant}, $\mathtt{AstroPy}$~\citep{Astropy2018}, $\mathtt{Sklearn}$~\citep{scikit-learn}, $\mathtt{AstroML}$~\citep{astroML}, $\mathtt{Dophot}$~\citep{Schechter1993, Alonso2012}, 
$\mathtt{VOSA}$~\citep{Bayo2008}, 
$\mathtt{STILTS}$~\citep{Taylor2006}. }

\begin{longrotatetable}
\begin{table}
\tabcolsep=2pt
\begin{tabular}{c c c c c c c c c c c c c c }
\hline\hline
&&&&&&&&&&&&&\\[-10pt]
Name              & $\alpha(2000)$     & $\delta(2000)$     &    & $[\alpha/{\rm H}]$ & $\log g$  & $T_{\rm eff}$ & RV   & $\pi$ (mas)     & $\mu_\alpha\cos\delta$        & $\mu_\delta$      & G   & BP-BR & Sp type \\
&&&&&&&&&&&&&\\[-10pt]
\hline 
&&&&&&&&&&&&&\\[-10pt]
VVVJ13081183-6317548 & 197.0492770 & -63.2985550 & 0.4501 & -0.0997   & 4.8207 & 4340 & -70.32 & 3.9893$\pm$0.1396 & -69.497$\pm$0.163 & -1.755$\pm$0.165  & 13.91 & 1.59 & K5-7  \\
VVVJ13081329-6242491 & 197.0553602 & -62.71362753 & 0.0518 & -0.0437   & 5.1028 & 3846 & 9.24  & 15.0873$\pm$0.0305 & -106.512$\pm$0.039 & -115.861$\pm$0.039 & 13.42 & 2.40 & M3-5  \\
2M13182865-6253076          & 199.6193776 & -62.88544519 & 0.5706 & -0.0994   & 5.1461 & 4951 & -17.55 & 5.3724$\pm$0.2526 & -80.335$\pm$0.346 & -26.341$\pm$0.36  & 12.74 & 1.285 & K3-5  \\ 
&&&&&&&&&&&&&\\[-10pt]
\hline
\end{tabular}
\caption{High proper motion stars observed by APOGEE.}
\label{high_pm}
\end{table}

\begin{table}\tiny
\tabcolsep=1.1pt
\begin{tabular}{l l l l l l l l l r l r r r r r r r l l l l l r r r r r r r r r r r r r r r r r r l}
\hline\hline
&&&&&&&&&&&&&&&&&&&&&&&&&&&&&&&&&&&&&&&&&\\[-5pt]
  \multicolumn{1}{c}{0ID} &
  \multicolumn{1}{c}{$\alpha(2000)$} &
  \multicolumn{1}{c}{$\delta(2000)$} &
  \multicolumn{1}{c}{$l$} &
  \multicolumn{1}{c}{$b$} &
  \multicolumn{1}{c}{Mean$K_{\rm S}$} &
  \multicolumn{1}{c}{$K_{\rm S}$err} &
  \multicolumn{1}{c}{Nobs} &
  \multicolumn{1}{c}{Ampl} &
  \multicolumn{1}{c}{$\eta$} &
  \multicolumn{1}{c}{TileID} &
  \multicolumn{1}{c}{$K_{\rm S}55264$} &
  \multicolumn{1}{c}{mag\_err4} &
  \multicolumn{1}{c}{$J$} &
  \multicolumn{1}{c}{$J$err} &
  \multicolumn{1}{c}{$H$} &
  \multicolumn{1}{c}{$H$err} &
  \multicolumn{1}{c}{$J-H$} &
  \multicolumn{1}{c}{$H-K_{\rm S}$} &
  \multicolumn{1}{c}{Class} &
  \multicolumn{1}{c}{Var Class} &
  \multicolumn{1}{c}{Comment} &
  \multicolumn{1}{c}{Reference} \\
  \multicolumn{1}{c}{[3.6]} &
  \multicolumn{1}{c}{[3.6]err} &
  \multicolumn{1}{c}{[4.5]} &
  \multicolumn{1}{c}{[4.5]err} &
  \multicolumn{1}{c}{[5.8]} &
  \multicolumn{1}{c}{[5.8]err} &
  \multicolumn{1}{c}{[8.0]} &
  \multicolumn{1}{c}{[8.0]err} &
  \multicolumn{1}{c}{[24]} &
  \multicolumn{1}{c}{[24]err} &
  \multicolumn{1}{c}{W1} &
  \multicolumn{1}{c}{W1err} &
  \multicolumn{1}{c}{W2} &
  \multicolumn{1}{c}{W2err} &
  \multicolumn{1}{c}{W3} &
  \multicolumn{1}{c}{W3err} &
  \multicolumn{1}{c}{W4} &
  \multicolumn{1}{c}{W4err} &
  \multicolumn{1}{c}{$\alpha$} &
  \multicolumn{1}{c}{Class($\alpha$)} \\
&&&&&&&&&&&&&&&&&&&&&&&&&&&&&&&&&&&&&&&&&\\[-5pt]
\hline
&&&&&&&&&&&&&&&&&&&&&&&&&&&&&&&&&&&&&&&&&\\[-5pt]
  d045\_v1 & 194.045408 & -63.311269 & 303.4645635 & -0.4444275 & 16.422 & 0.032 & 42 & 1.444 & 0.414 & d045 &  &  &  &  &  &  &  &  &  & LAE &  &  \\ 14.051 & 0.099 & 13.213 & 0.095 & 12.326 & 0.143 & 11.541 & 0.063 & 7.87 & 0.1 &  &  &  &  &  &  &  &  & 0.459 & Class I\\
  &&&&&&&&&&&&&&&&&&&&&&&&&&&&&&&&&&&&&&&&&\\[-5pt]
    d045\_v2 & 194.082928 & -63.608539 & 303.4757633 & -0.7419591 & 12.348 & 0.017 & 51 & 0.742 & 0.157 & d045 & 11.889 & 0.058 & 13.659 & 0.121 & 12.616 & 0.098 & 1.043 & 0.727 & T & Dipper &  &  \\ 11.176 & 0.075 & 10.888 & 0.073 & 10.551 & 0.089 & 10.378 & 0.074 & 5.51 & 0.07 & 10.766 & 0.03 & 10.49 & 0.027 & 8.738 & 0.184 & 4.854 & 0.104 & 0.176 & Flat\\
      &&&&&&&&&&&&&&&&&&&&&&&&&&&&&&&&&&&&&&&&&\\[-5pt]
   d045\_v3 & 194.089278 & -63.331243 & 303.4838829 & -0.4647675 & 16.027 & 0.022 & 52 & 1.08 & 0.285 & d045 & 15.663 & 0.059 &  &  & 17.522 & 0.067 &  & 1.859 &  & Fader & in-group &  \\ 12.548 & 0.075 & 11.149 & 0.057 & 10.346 & 0.148 &  &  &  &  &  &  &  &  &  &  &  &  &  & \\
     &&&&&&&&&&&&&&&&&&&&&&&&&&&&&&&&&&&&&&&&&\\[-5pt]
  d045\_v4 & 194.101456 & -63.39115 & 303.4881923 & -0.5247683 & 13.912 & 0.014 & 45 & 0.686 & 0.553 & d045 & 13.807 & 0.058 &  &  & 17.063 & 0.066 &  & 3.256 &  & LAE &  & Robitaille \\ 9.91 & 0.038 & 8.612 & 0.038 & 7.376 & 0.024 & 6.773 & 0.022 &  &  & 10.161 & 0.023 & 8.255 & 0.019 & 5.98 & 0.025 & 3.652 & 0.024 & -0.043 & Flat\\
    &&&&&&&&&&&&&&&&&&&&&&&&&&&&&&&&&&&&&&&&&\\[-5pt]
   d045\_v5 & 194.242266 & -63.77788 & 303.542943 & -0.9126947 & 14.142 & 0.019 & 50 & 1.038 & 0.319 & d045 & 13.576 & 0.058 & 15.01 & 0.12 & 14.154 & 0.098 & 0.856 & 0.578 & T & Fader &  & \\ 13.226 & 0.079 & 12.787 & 0.071 & 12.723 & 0.249 & 12.315 & 0.121 &  &  &  &  &  &  &  &  &  &  & -1.326 & Class II\\
     &&&&&&&&&&&&&&&&&&&&&&&&&&&&&&&&&&&&&&&&&\\[-5pt]
     d045\_v6 & 194.406342 & -63.158305 & 303.6303178 & -0.2949481 & 12.858 & 0.023 & 51 & 1.049 & 0.458 & d045 & 12.432 & 0.057 & 14.098 & 0.12 & 13.134 & 0.098 & 0.964 & 0.702 & T & Eruptive & FIRE &  \\ 12.9 & 0.036 & 12.707 & 0.061 & 12.274 & 0.104 & 12.225 & 0.203 &  &  & 11.75 & 0.031 & 11.379 & 0.027 & 9.723 & 0.061 & 8.084 &  & -0.239 & Flat\\
       &&&&&&&&&&&&&&&&&&&&&&&&&&&&&&&&&&&&&&&&&\\[-5pt]
  d045\_v7 & 194.475246 & -63.443083 & 303.654274 & -0.5804006 & 14.287 & 0.015 & 53 & 0.8 & 0.447 & d045 & 14.044 & 0.058 & 15.276 & 0.12 & 14.57 & 0.098 & 0.706 & 0.526 & T & Dipper &  &  \\ 13.293 & 0.064 & 12.937 & 0.121 & 12.996 & 0.329 & 12.562 & 0.301 &  &  &  &  &  &  &  &  &  &  & -1.621 & Class III\\
    &&&&&&&&&&&&&&&&&&&&&&&&&&&&&&&&&&&&&&&&&\\[-5pt]
  d045\_v8 & 194.502338 & -63.006601 & 303.6775207 & -0.1443672 & 13.714 & 0.026 & 46 & 1.265 & 0.609 & d045 & 13.252 & 0.057 & 15.783 & 0.12 & 14.295 & 0.098 & 1.488 & 1.043 & T & Eruptive &  &  \\ 11.906 & 0.042 & 11.203 & 0.034 & 10.564 & 0.044 & 9.666 & 0.022 &  &  &  &  &  &  &  &  &  &  & 0.854 & Class I\\
    &&&&&&&&&&&&&&&&&&&&&&&&&&&&&&&&&&&&&&&&&\\[-5pt]
  d045\_v9 & 194.502973 & -62.98427 & 303.6783789 & -0.1220508 & 14.545 & 0.013 & 46 & 0.766 & 1.053 & d045 & 14.204 & 0.057 & 15.593 & 0.12 & 14.643 & 0.098 & 0.95 & 0.439 & F & STV &  &  \\ 13.457 & 0.046 & 12.9 & 0.065 & 12.349 & 0.107 & 11.514 & 0.05 &  &  &  &  &  &  &  &  &  &  & 0.366 & Class I\\
    &&&&&&&&&&&&&&&&&&&&&&&&&&&&&&&&&&&&&&&&&\\[-5pt]
   d045\_v10 & 194.558735 & -62.93687 & 303.7049504 & -0.0753248 & 16.052 & 0.027 & 47 & 1.405 & 0.698 & d045 &  &  &  &  &  &  &  &  &  & STV &  & Robitaille \\ 11.691 & 0.042 & 10.617 & 0.053 & 9.583 & 0.028 & 8.717 & 0.026 &  &  & 12.576 & 0.039 & 10.578 & 0.022 & 8.08 & 0.026 & 5.595 & 0.04 & -0.007 & Flat\\
   &&&&&&&&&&&&&&&&&&&&&&&&&&&&&&&&&&&&&&&&&\\[-5pt]
    \multicolumn{42}{l}{... And 186 more objects ...} \\
    &&&&&&&&&&&&&&&&&&&&&&&&&&&&&&&&&&&&&&&&&\\[-5pt]
\hline
\end{tabular}
\caption{Parameters of the VVV variables considered in this work. Description of each column and the full version of the table is available at the CDS.
}
\label{v4_table}
\end{table}
\end{longrotatetable}
\bibliography{references} 


\end{document}